\documentclass[usenatbib]{mnras}
\usepackage{graphicx}

\usepackage{multirow}
\usepackage{bigdelim}

\title[New Herbig-Haro objects and outflows in Mon R1 association]{New Herbig-Haro objects and outflows in Mon R1 association}
\author[T.A. Movsessian, T. Yu. Magakian, and S. N. Dodonov]{T.A. Movsessian,$^{1}$\thanks{E-mail:tigmov@bao.sci.am}, T. Yu. Magakian $^{1}$\thanks{E-mail:tigmag@sci.am}, S. N. Dodonov,$^{2}$\thanks{E-mail:dodo@sao.ru} 
\footnotemark[1]\thanks{This work is  based on observations conducted with the 1-m Schmidt telescope of the Byurakan Astrophysical Observatory of the National Academy of Sciences of Armenia.}\\
$^{1}$Byurakan Astrophysical Observatory,  Aragatsotn reg.,Armenia\\
$^{2}$Special Astrophysical Observatory, N.Arkhyz, Karachaevo-Cherkesia, 369167 Russia}
\usepackage{amssymb}

\begin{document}

\date{}

\pagerange{\pageref{firstpage}--\pageref{lastpage}} \pubyear{2002}

  \maketitle

\label{firstpage}

\begin{abstract}

We present results of the narrow-band H$\alpha$ and [S~\textsc{ii}] imaging survey of Mon R1 association,
performed with the 1 m Schmidt telescope of the Byurakan Observatory. Our observations covered one degree
field near the center of the association. As a result of this study twenty new Herbig-Haro knots were
discovered, some of which form collimated outflows. Among the most extended ones are HH 1203 and HH 1196, which have a length near one parsec or even more.
In the course of search for the probable sources of HH objects several new nebulous stars were found. A list of all nebulous stellar objects in the Mon R1 area   under study is  presented, with the detailed description of most interesting objects. The near infrared data from the GLIMPSE360 and WISE surveys allowed to find several more objects, related to  Mon R1, some of them with optical counterparts, as well as to outline at least three probable H$_2$\ collimated flows
from the deeply embedded pre-main-sequence objects.
The probable members of Mon R1 were selected by their distances, their bolometric luminosities and extinctions were estimated. Among the outflow sources three embedded objects with luminosities greater than 10 L$_{\sun}$ were found. The mean distance to Mon R1 complex is estimated as 715 pc.  

\end{abstract}


\section{Introduction}
Herbig-Haro objects represent shocked excitation zones where supersonic flows from young stellar objects (YSO) collide with interstellar medium and form small cloudlets with pure emission spectrum including permitted and low excitation forbidden emission lines 
([O~\textsc{i}], [S~\textsc{ii}] etc). For a long time they have been recognised as a sign of high  and recent activity of star formation in molecular clouds. The discovering of new HH objects is important for several reasons. First, HH objects are important in themselves, but especially because
they provide a historical record of the sources outburst activity. By
accumulating as many outflow histories, especially by discovering giant HH
flows, we learn about the history of young stars on timescales hundreds of
times a human lifetime. Second,  newly found HH groups and flows help to detect  new, previously unknown star forming regions as well as to better understand the structure and the formation of the stars of various masses in large clouds.

\begin{figure*}
 \includegraphics[width=400pt]{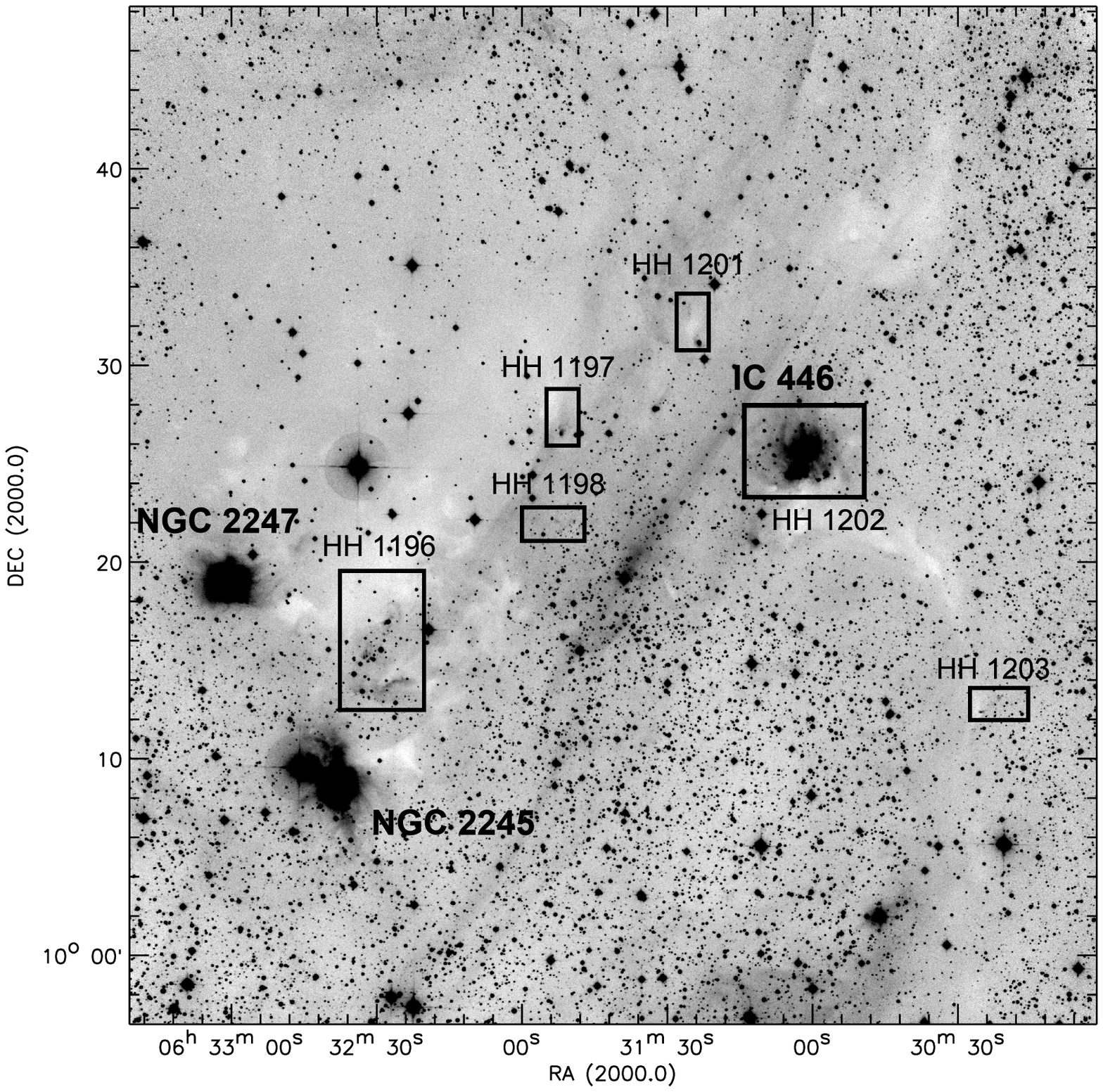}
 \caption{Whole observed field in H$\alpha+$[S~\textsc{ii}] emission, built from the images, obtained with 1 m telescope.  The areas including newly discovered HH objects and outflows, are shown by rectangles. Also three major reflection nebulae are marked.  }
\label{fig1}
\end{figure*} 
 
We continue our searches for HH objects in the dark clouds, started in Byurakan observatory more than twenty years ago, with the new equipment and wider-field telescope (see sec.2). Among the first targets we included  in our program the Mon R1 association \citep{Racine}. 

This group of the stars, illuminating several bright reflection nebulae of various sizes, is located to the north-west from the famous Mon~OB1 association. Some of these stars, e.g. those associated with  NGC~2245 and NGC~2247, are well-known and studied by many authors. In general, Mon~R1 contains at least about 30  young stellar objects (YSOs), found in optical range \citep{Herbst}.  The distance of Mon~R1 usually is assumed to be 800 pc, same as for the nearby Mon~OB1
\citep[see][for detailed review]{dahm}.
One can see that YSOs in  Mon~R1 are spatially divided into two groups; one includes NGC~2245, NGC~2247 
and IC~446, and another one is projected on the large reflection nebula IC~2169. Recent observations indicate the significant star forming activity in the later group \citep{Bhadari}.

In contrast with the adjacent Mon~OB1, Mon~R1 received much lower attention. Nearly all observational studies were concentrated on IC~446, where a compact group of emission-line stars, including HAeBe star VY~Mon,  was found by \citet{ck}. Further  observations in the  infrared range confirmed an existence of the small cluster of probable YSOs around VY~Mon \citep[and references therein]{WL,gutermuth}. No searches of HH objects in the Mon~R1 field were performed so far.

 Our field includes three brightest nebulae NGC~2245, NGC~2247
and IC~446, thus  making our target the dense core of Mon~R1 association.

\section[]{Observations}
The images were obtained on the nights of Feb. 3--4 2019  with 1m Schmidt telescope of Byurakan observatory, which was upgraded during 2013--2015 and equipped with CCD detector. Reworked 4K$\times$4K Apogee (USA) liquid-cooled CCD camera was used as a detector with a pixel size of 0.868\arcsec\ and field of view of about 1 square degree \citep{Dodo}. 

Narrow-band filters centered on 6560 \AA\ and 6760 \AA, both with a
FWHM of 100 \AA, were used to obtain H$\alpha$ and [S~\textsc{ii}] images, respectively. A midband filter, centered on 7500 \AA\ with a FWHM of 250 \AA,\  was used for the continuum imaging.

A dithered set of 5 min exposures was obtained in each filter. Effective exposure time in H$\alpha$ equaled 6000 sec, in [S~\textsc{ii}] -- 7200 sec and in continuum -- 2400 sec. Images were reduced in the standard manner using IDL package developed by one of authors (SND), which includes bias subtraction, cosmic ray removal, and flat fielding using ``superflat field'', constructed by several images.

The search of HH objects was done by classic technique, suggested in 1975 by \citet{VdB}. Present approach includes the comparison of
H$\alpha$, [S~\textsc{ii}] and I continuum images. As the practice shows, this is sufficient for the
reliable identification of HH objects in overwhelming majority of cases.

\begin{figure*}
        \centering
        \begin{tabular}{@{}ccc@{}}
                \includegraphics[width=125pt]{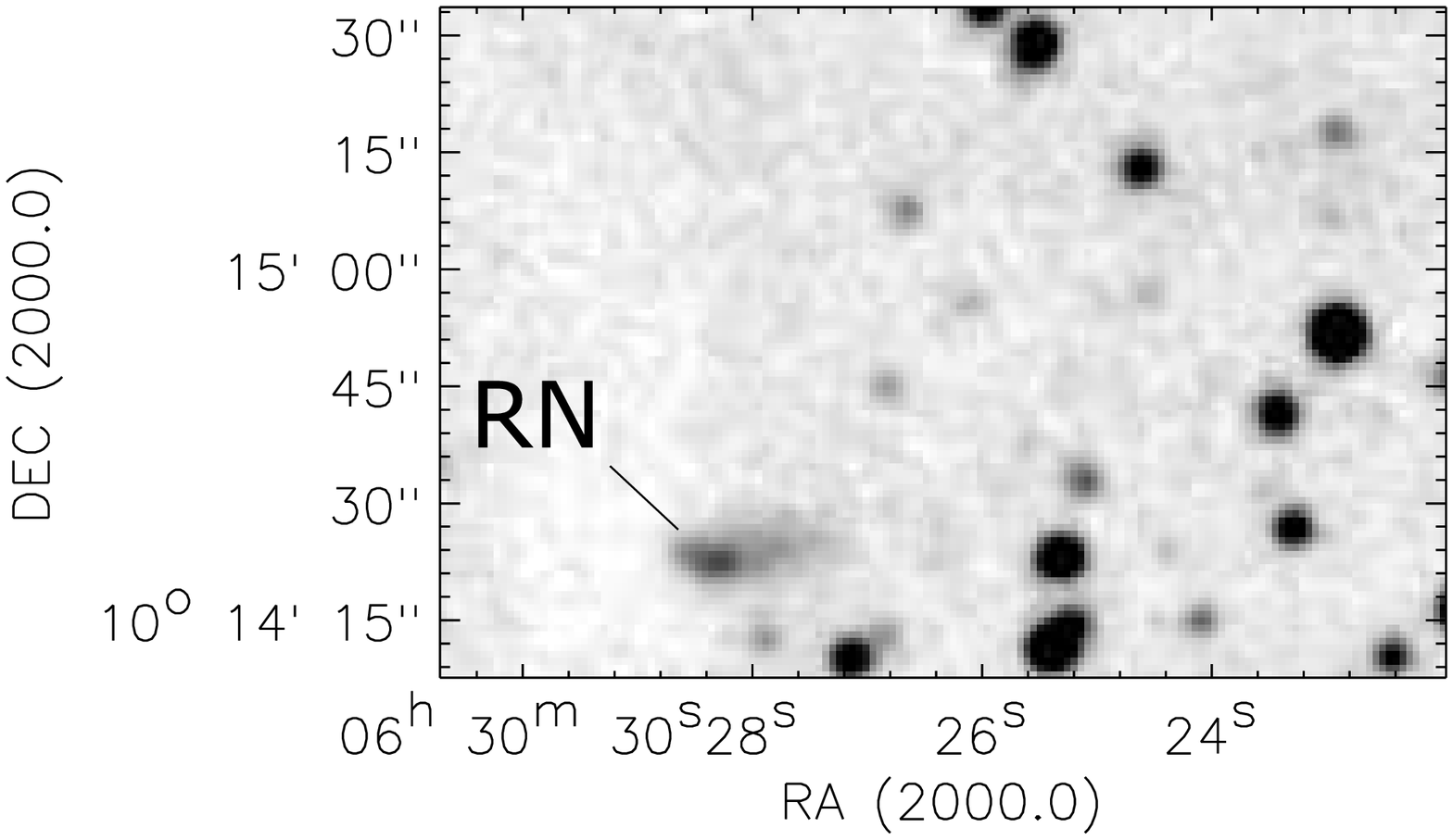}&
                \includegraphics[width=125pt]{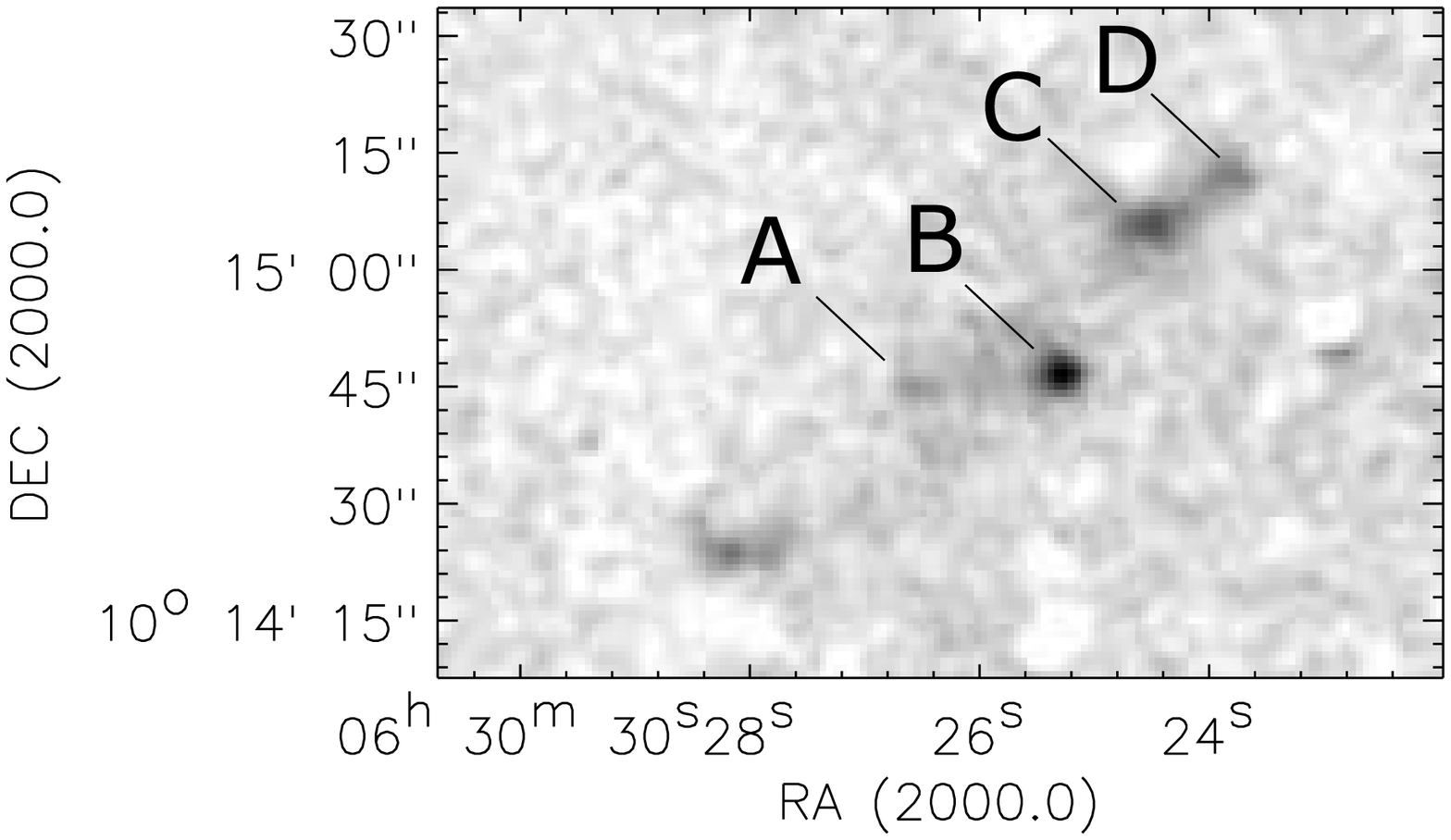}&
                \includegraphics[width=125pt]{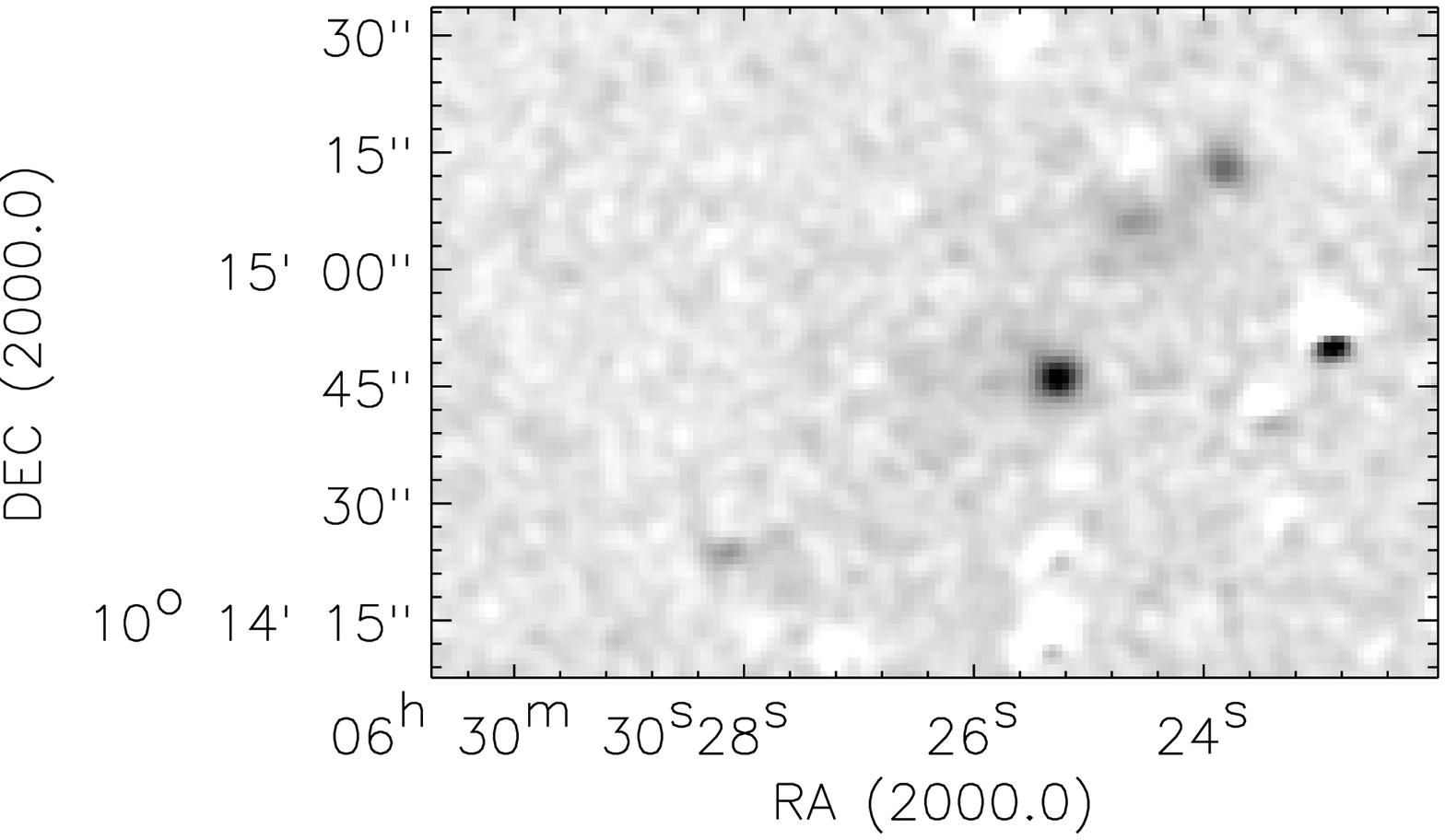}
        \end{tabular}
        \caption{Images of HH~1203 outflow system: continuum image (left panel), continuum
subtracted H$\alpha$ (central panel) and continuum subtracted [S~\textsc{ii}] (right panel).}
        \label{HH1203}
\end{figure*}

\section{Results}

 The area under study includes  three bright reflection nebulae (NGC~2245, NGC~2247 and IC~446), as well as several isolated dark lanes and emission-reflection filaments, which are stretched in SE-NW direction through the whole field.
A considerable amount of totally new HH knots and groups, standing out by their pure emission-line spectrum, were found in this area.

On the other hand, in the course of HH knots search, we noted several nebulous stars, which can be the probable members of Mon~R1. Some of them can represent exciting sources of HH objects. All these objects are described and discussed in the following sections.

\subsection{HH objects and flows}

\begin{table*}
 \centering
 \begin{minipage}{140mm}
  \caption{The coordinates of HH objects and HH flows in the Mon~R1 field.}
  \begin{tabular}{llll@{}}
  \hline
   Name     &  RA(2000)    & Decl.(2000) & Notes  \\
 \hline
  HH 1203A & 06 30 26.8 & $+$10 14 49& \rdelim\}{4}{8cm}[Collimated HH flow, associated with a cometary nebula] \\
  HH 1203B & 06 30 25.4 & $+$10 14 50 \\
  HH 1203C & 06 30 24.7 & $+$10 15 11 \\
  HH 1203D & 06 30 24.0 & $+$10 15 17 \\ 

 \hline
  HH 1202A & 06 31 12.8 & $+$10 25 35 & \rdelim\}{3}{8cm}[HH objects in the VY~Mon region] \\ 
  HH 1202B & 06 31 09.0 & $+$10 27 03 \\ 
  HH 1202C & 06 31 08.3 & $+$10 26 26 \\ 
 \hline
  HH 1201A & 06 31 32.5 & $+$10 34 22 &  \rdelim\}{2}{8cm}[A pair of HH knots about 2\arcmin\ north from the LkH$\alpha\ $342] \\ 
  HH 1201B & 06 31 32.5 & $+$10  34 01 \\
 \hline 
  HH 1198A & 06 32 02.4 & $+$10 23 13   &  \rdelim\}{4}{8cm}[An isolated group of HH knots]    \\
  HH 1198B & 06 31 57.8 & $+$10 23 07   &  \\
  HH 1198C & 06 31 56.9 & $+$10 22 39   &  \\
  HH 1198D & 06 31 54.2 & $+$10 23 09   &  \\
 \hline
  HH 1197  & 06 31 58.8 & $+$10 27 45   & Compact HH object near RNO~72 \\
 \hline
  HH 1196A & 06 32 30.9 & $+$10 17 48   & \rdelim\}{6}{8cm}[A well-defined flow from embedded infrared source] \\
  HH 1196B & 06 32 32.3 & $+$10 16 58   & \\
  HH 1196C & 06 32 35.1 & $+$10 16 22   & \\
  HH 1196D & 06 32 35.9 & $+$10 16 04   & \\
  HH 1196E & 06 32 36.2 & $+$10 15 43   & \\
  HH 1196F & 06 32 37.9 & $+$10 13 56   & \\
 \hline

\hline  
\end{tabular}
\end{minipage}
\end{table*}

 A H$\alpha+$[S~\textsc{ii}] image, which is covering one square degree field of the Mon~R1 region, is shown on Fig.\ref{fig1}. The zones with newly discovered HH objects and HH flows are marked by rectangles. All HH objects found are listed in Table 1 in order of their right ascension.

\subsubsection{HH~1203 flow}
This HH flow includes four HH objects in a chain, embedded in the general diffuse emission (marked in Fig.\,\ref{HH1203} as knots A, B, C and D).  Brightest knot (B) in this chain is somewhat displaced from the axis of whole complex.

\begin{figure}
        \includegraphics[width=210pt]{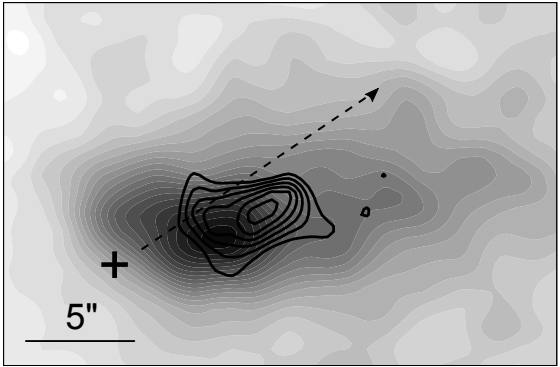}
        \caption{Image of the reflection nebula near HH 1203 in continuum (gray scale) and in 
        [S~\textsc{ii}] emission (isolines). The position of IRAS~06277+1016 infrared source is marked by the cross. The allWISE J063028.65+101425.4 source is shifted to 2.1\arcsec\ in p.a.=32\degr\ direction from this position. Short emission jet, pointing in the north-west direction along the axis of HH flow (shown by dashed line), is well seen.}
        \label{HH1203Neb}
\end{figure}

To the south-east from this HH group there is a compact cone-shaped reflection nebula, which lies near the axis of this elongated group. On Fig.\,\ref{HH1203Neb} the continuum image of this reflection nebula with overlapped 
[S~\textsc{ii}] image is presented. This combination reveals the short emission jet, elongated in the direction of the axis of HH~1203 flow. Its existence   confirms the assumption that the star, obviously embedded into the nebula, is the source of HH 1203.

\subsubsection{HH objects in the area of IC~446 and VY~Mon}

Reflection nebula IC~446 is located near the center of Mon R1 association, on the edge  of a small dark cloud. As was mentioned above, it is surrounded by a cluster of emission-line and infrared stars, with dominating YSO VY~Mon \citep{casey}. 

\begin{figure*}
\centering
\begin{tabular}{@{}ccc@{}}
\includegraphics[width=120pt]{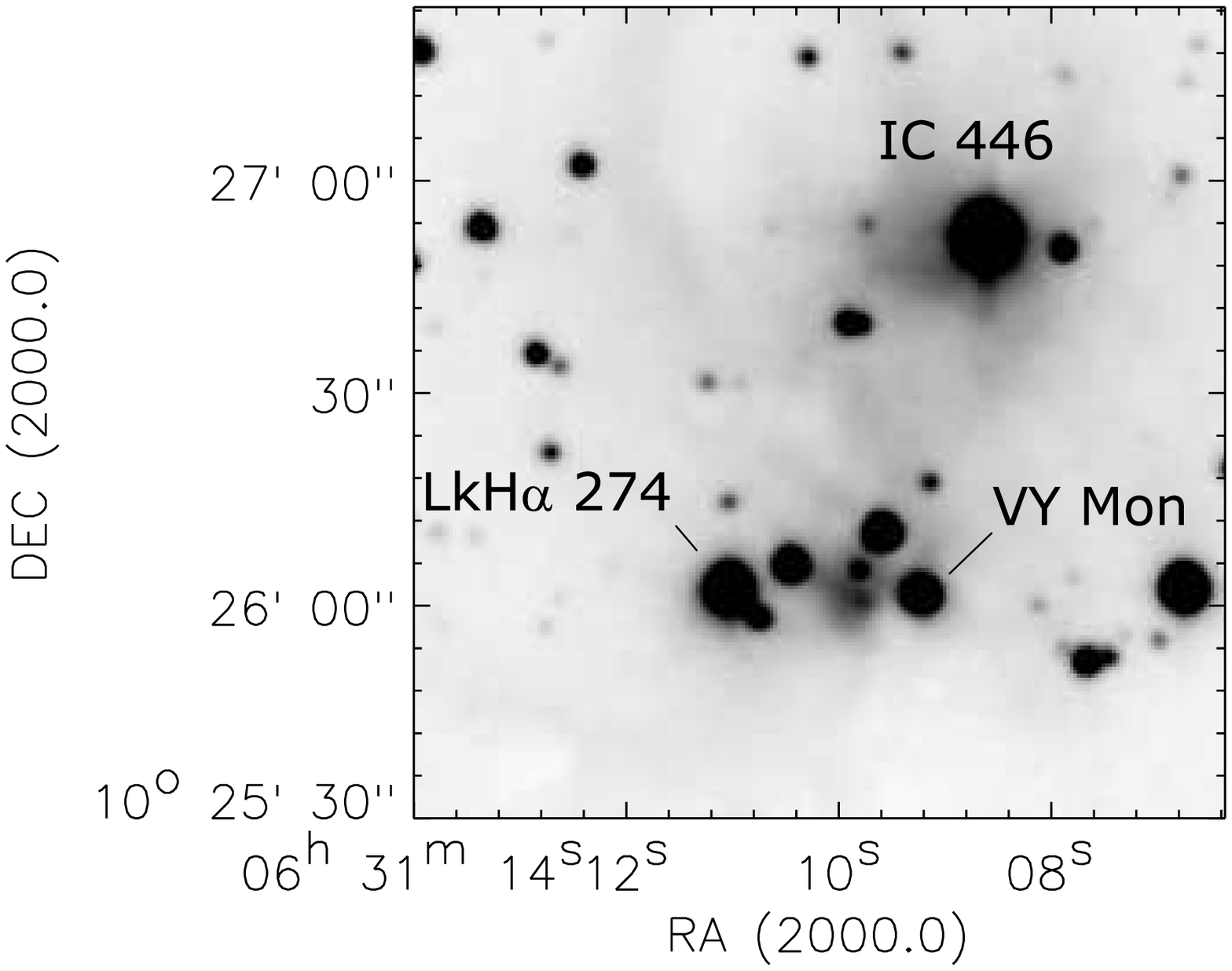} &
\includegraphics[width=120pt]{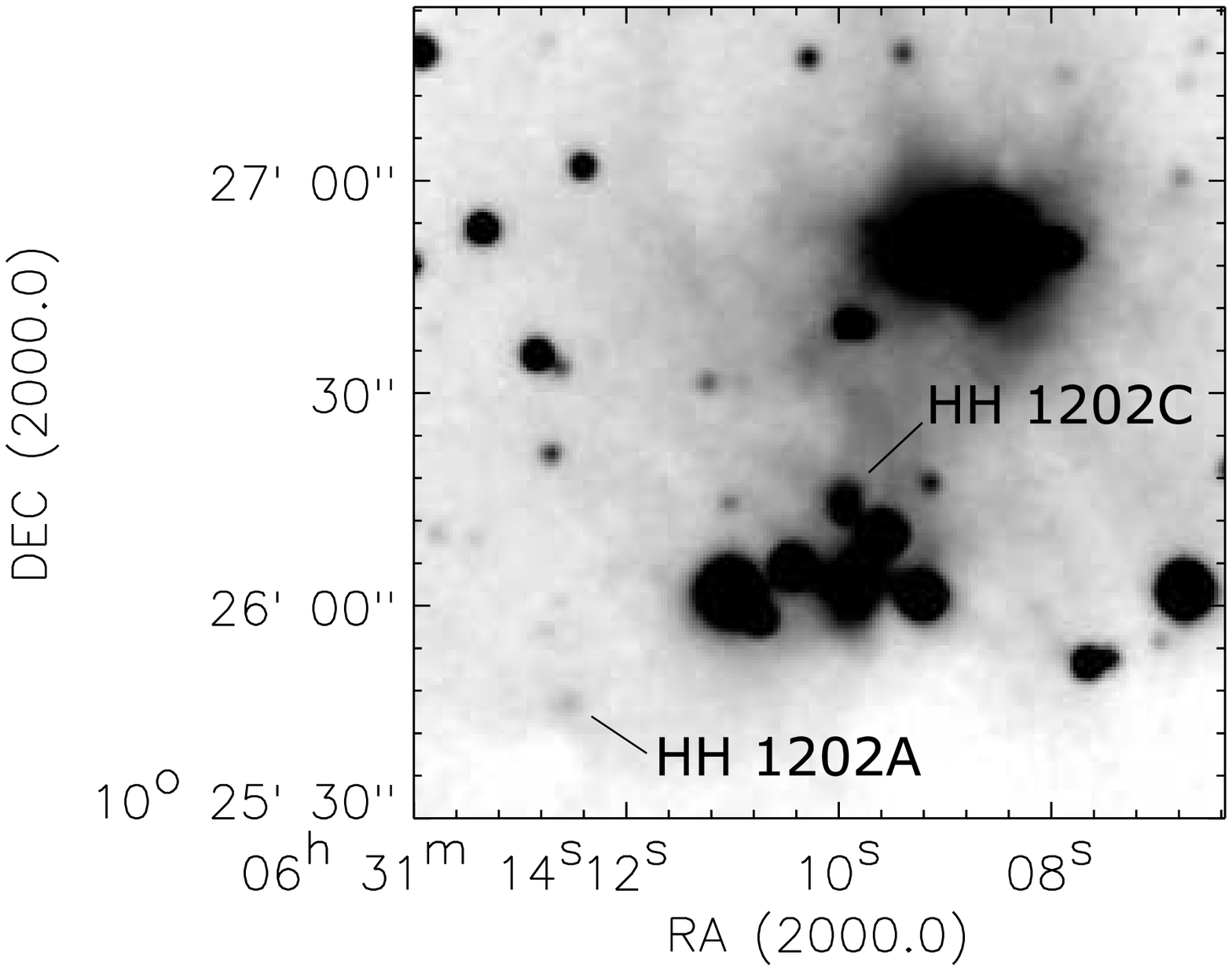} &
\includegraphics[width=118pt]{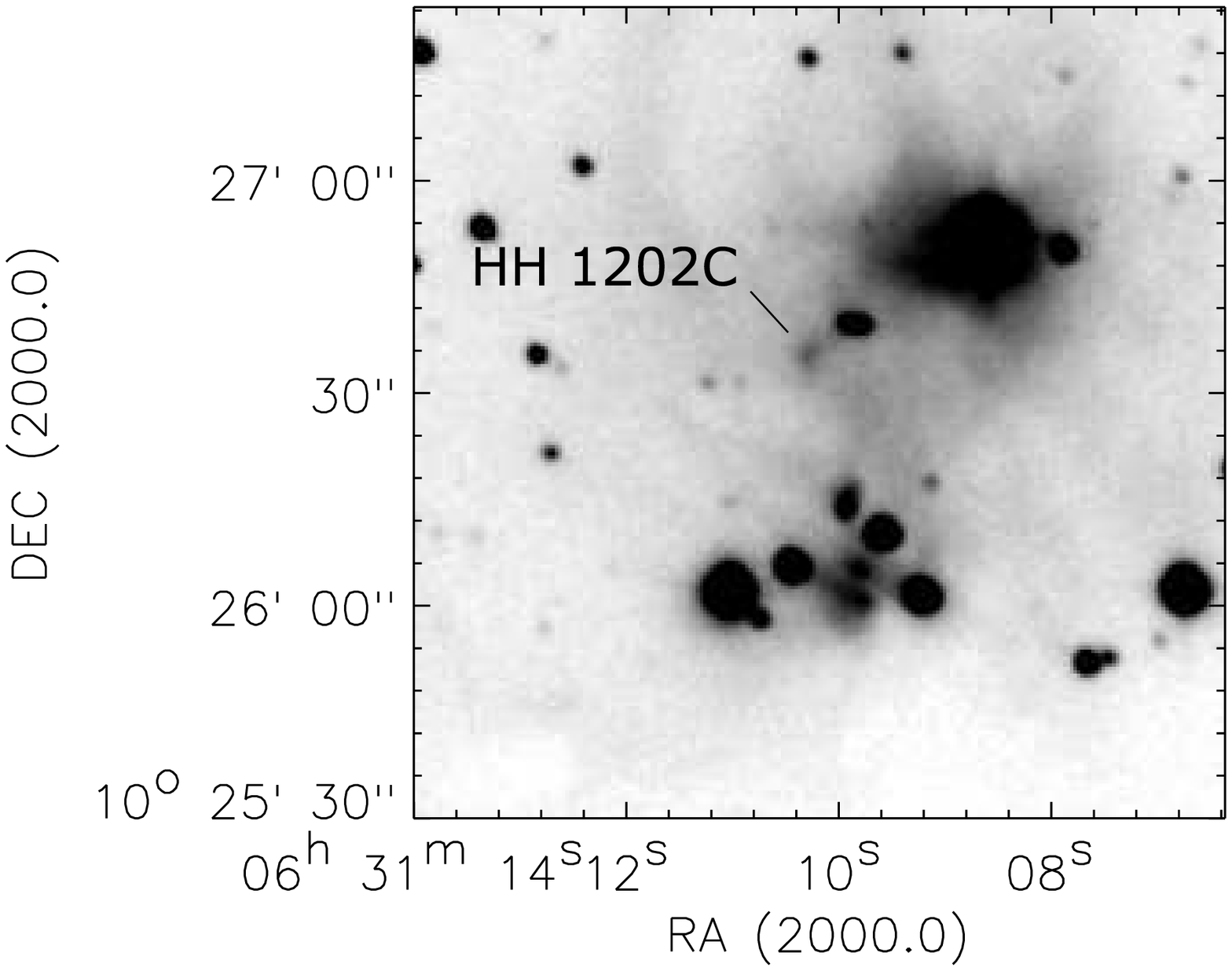}
\end{tabular}
\caption{Images of VY Mon field, including knots of HH 1202, in continuum (left panel), H$\alpha$ (central panel) and [S~\textsc{ii}] (right panel).}
\label{VYMon}
\end{figure*} 

We discovered three HH objects in this area  (Fig.\,\ref{VYMon}).  The brightest one (HH~1202C) is located about 30$\arcsec$ north-east from VY~Mon and is visible even  on DSS-2 red images. This object has comparatively equal brightness in [S~\textsc{ii}] and H$\alpha$. The second knot (HH~1202B) was detected in [S~\textsc{ii}] images and is not visible in H$\alpha$. It is located in $\approx$12$\arcsec$ to south-east from a close pair of stars, which are listed as 2MASS sources 06310840+1027111 and 06310815+1027109. The third one (HH~1202A) was found near the edge of dark nebula in $\approx$100\arcsec\ to the east from VY~Mon. This HH object is visible only in H$\alpha$ image. 

It is naturally to expect that VY~Mon, being the most luminous YSO in this field, can be a probable outflow source. Considering this it should be noted that the line, connecting HH~1202B and HH~1202C, passes near VY~Mon. Moreover, a faint cone-shaped reflection nebula with the axis oriented toward these two HH objects, can be traced near this very active star.  

The source of HH~1202A is not so obvious, but, probably, it can be associated with one of YSOs belonging to the cluster around IC~446.

\subsubsection{HH~1201}
\begin{figure}
        \centering
        \begin{tabular}{@{}cc@{}}
                \includegraphics[width=97pt]{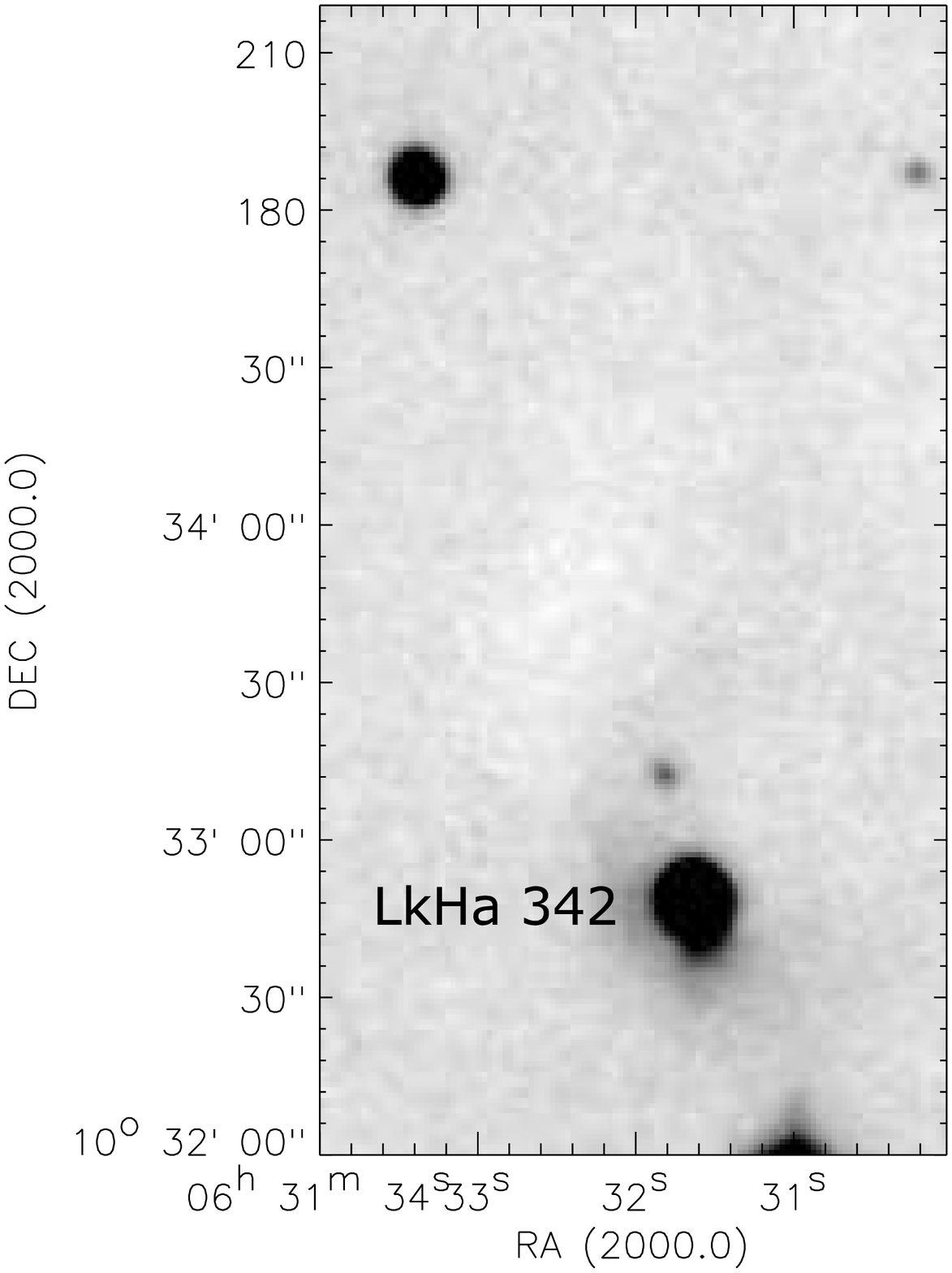}&
                \includegraphics[width=97pt]{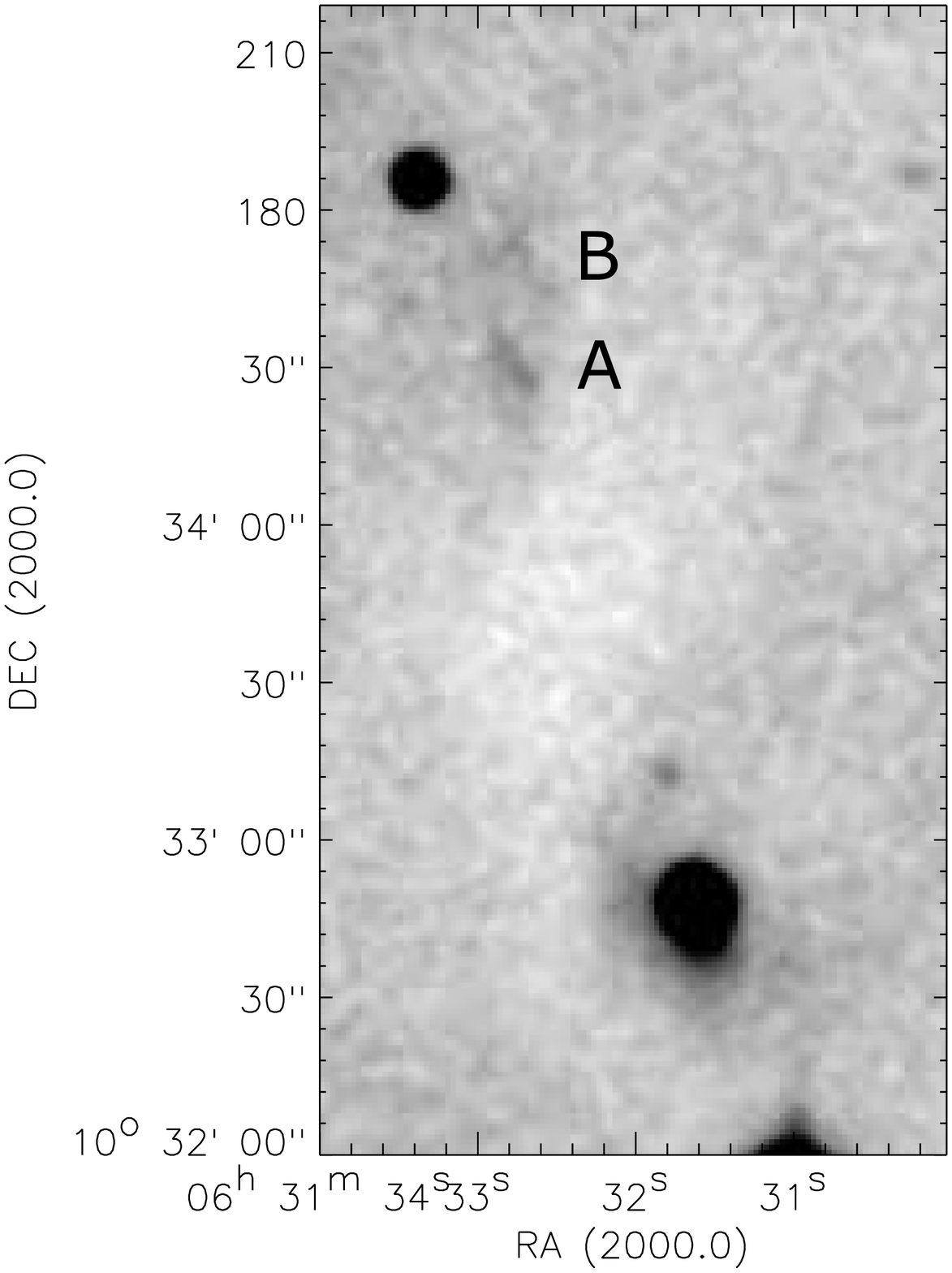}
        \end{tabular}
        \caption{Images of HH 1201 outflow: continuum image (left panel), H$\alpha$ image (right panel).}
        \label{HH1201}
\end{figure}

To the north-east from IC~446 a pair of HH knots was discovered. This group is elongated in north-south direction; knots are separated by about 20\arcsec\ (Fig.\,\ref{HH1201}). This object was detected in H$\alpha$ and is barely visible in [S~\textsc{ii}]. To the south from these objects young star LkH$\alpha$~342  associated with small optical reflection nebula can be found. This star is visible in 2MASS and WISE surveys, but it is not a prominent infrared source. The line, drawn along the oblong knot HH 1201 A,  approximately is directed toward LkH$\alpha$~342;   however, this does not exclude other possible candidates  for HH 1201 source.\ 

\subsubsection{HH~1198}
\begin{figure}
        \includegraphics[width=230pt]{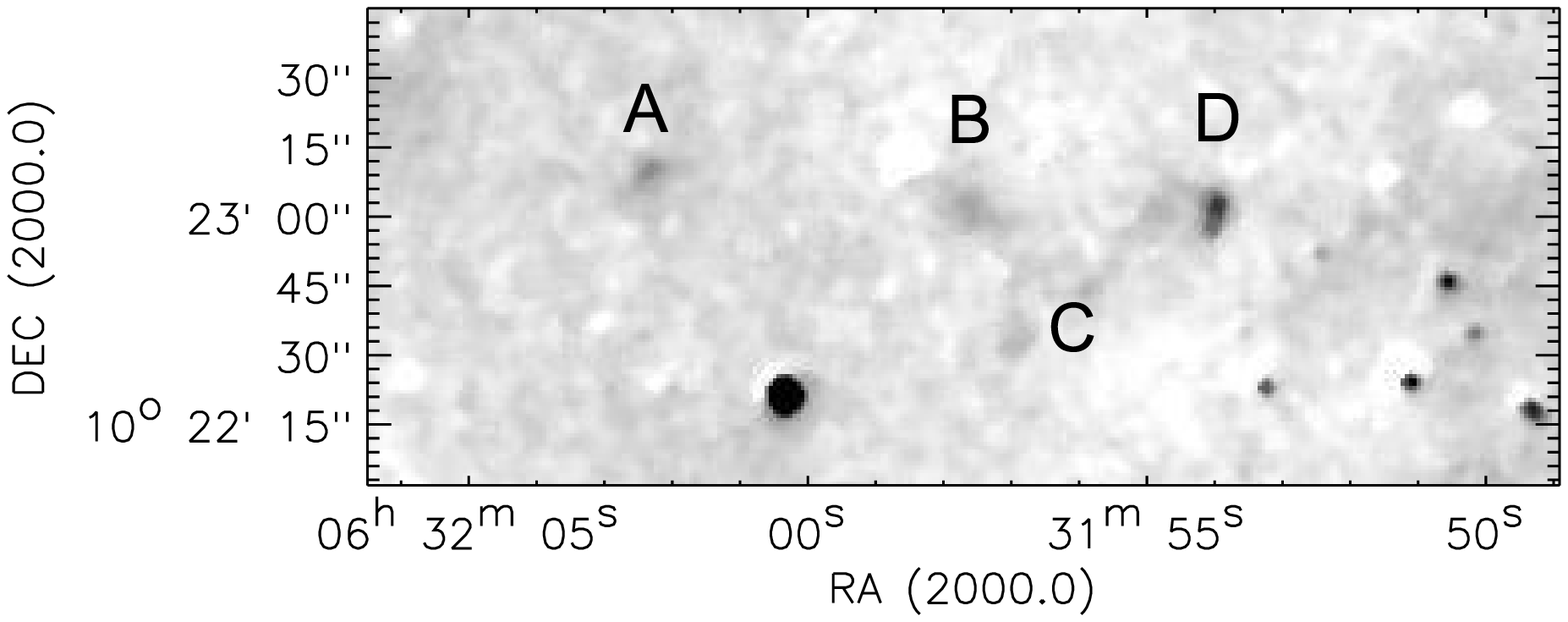}
        \caption{Continuum subtracted H$\alpha+$[S~\textsc{ii}] image of HH~1198.}
        \label{HH1198}
\end{figure}

This  HH group consists of four knots and is extended in east-west direction, with total length of about 2\arcmin \ (Fig.\,\ref{HH1198}). Knots A, B and D are located on the same line; knot C is somewhat aside and is connected with B and D by narrow filaments. As whole, the morphology of this group is vague  and resembles a heavily disrupted working surface of a flow,  source of which can be faraway in the uncertain direction.    

\subsubsection{HH~1197}
\begin{figure}
        \centering
        \begin{tabular}{@{}cc @{}}
                \includegraphics[width=114pt]{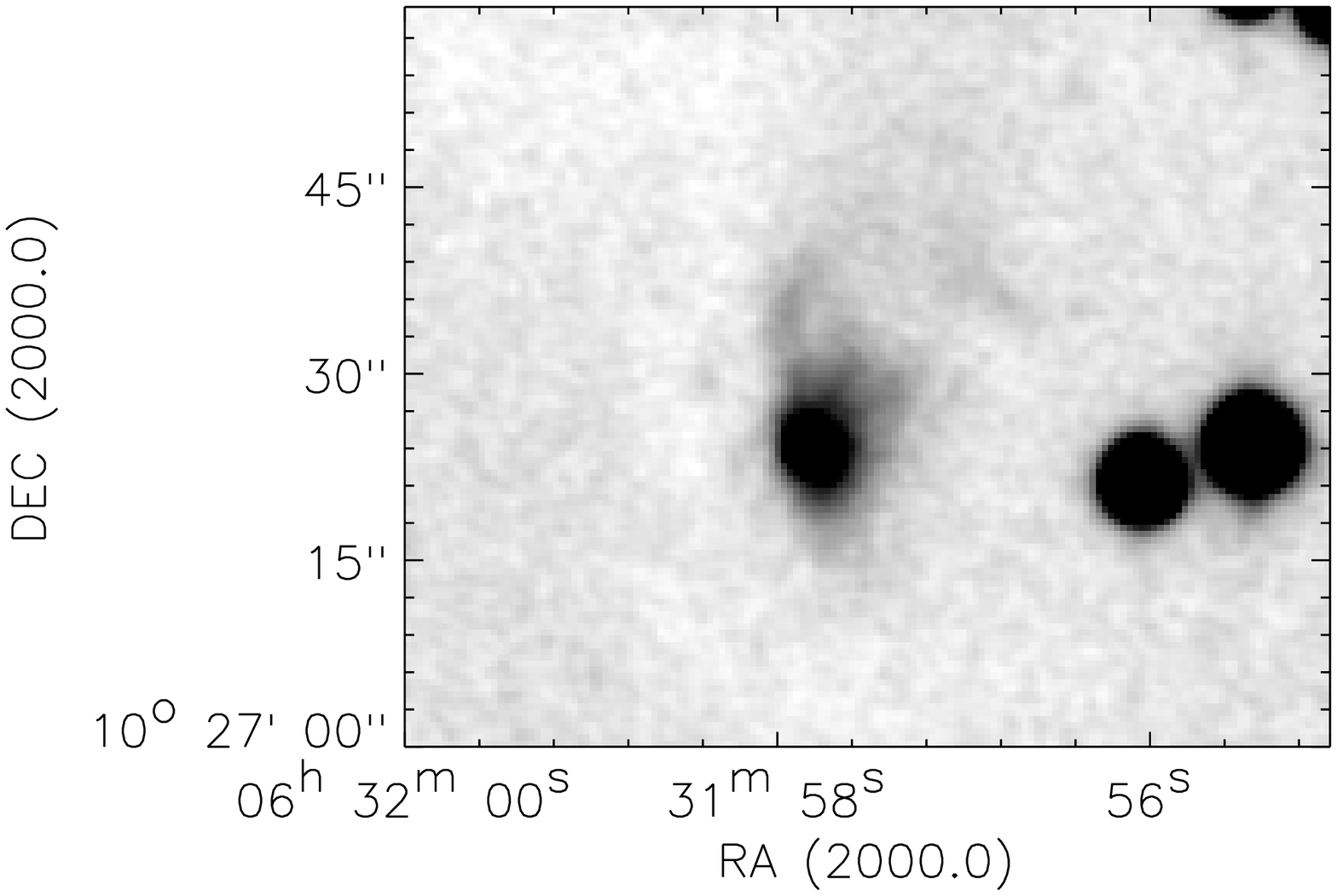}&
                \includegraphics[width=114pt]{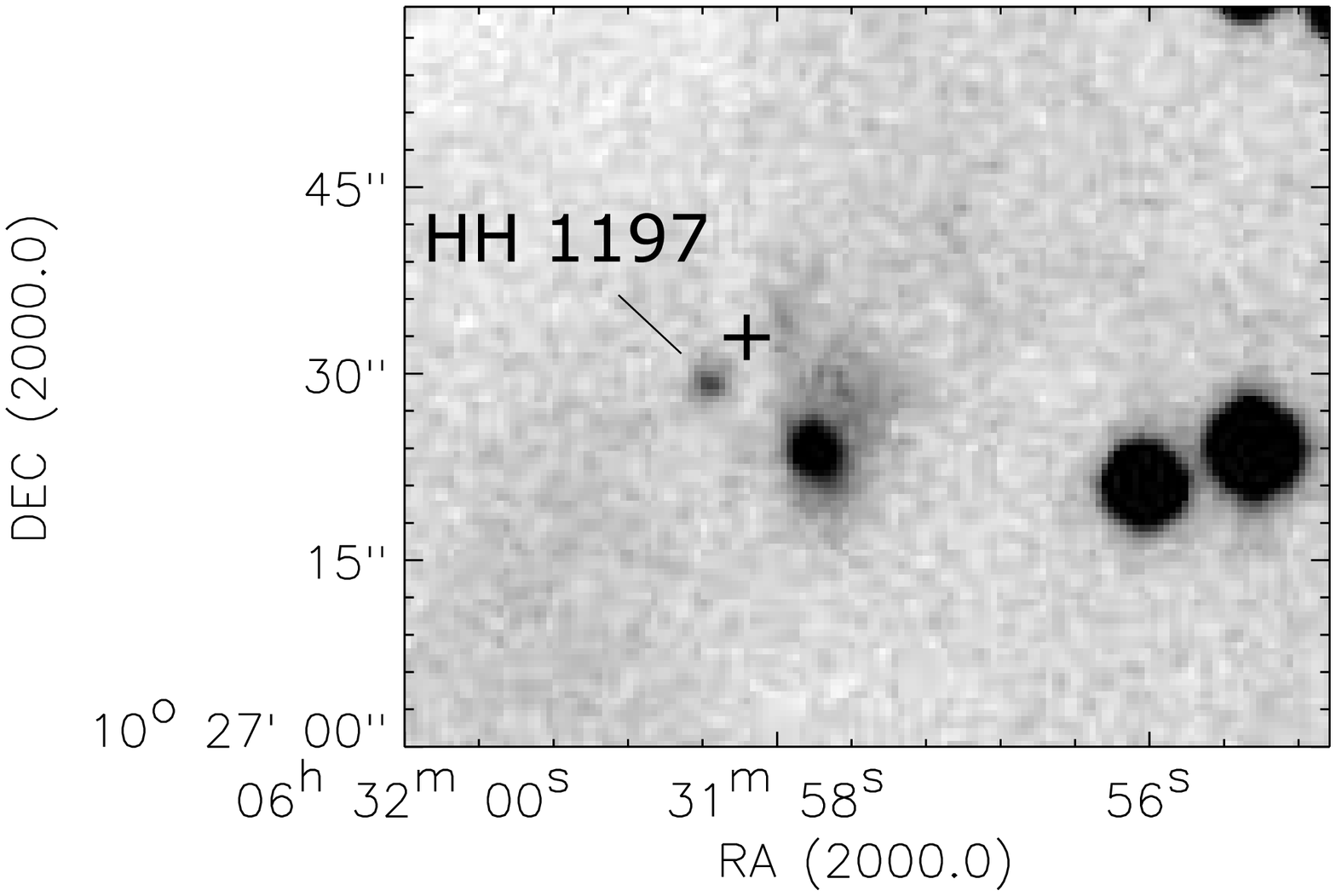}
        \end{tabular}
        \caption{Images of HH~1197 and nearby RNO~72 nebula in continuum (left panel) and in 
                [S~\textsc{ii}] emission line. Position of the 2MASS J06315870+1027474 infrared source is marked by a cross.}
        \label{HH1197}
\end{figure}

This compact HH object (Fig.\,\ref{HH1197}) is located near the small reflection nebula (RNO~72, HHL~43A), which will be discussed in the next section. This object is visible mainly in [S~\textsc{ii}] and is very faint in H$\alpha$, which is typical for HH jets, because they often have lower excitation than the HH objects representing working surfaces \citep{RB}; however, its traces also can be found in the continuum image. An infrared 2MASS source J06315870+1027474 (also IRAS 06292+1029) is very close to this object, being located $\approx$ 3\arcsec \ to the northwest. On the WISE 22 $\mu$m images it is rather bright. 

This object resembles the central object of HH~588 flow, a compact HH object with very low excitation, located near the IRAS 21388+5622 source, which probably represents the very short jet from the exciting star \citep{ogura,Movsessian2012}. Thus, one can assume that HH~1197 apparently propagates toward the southeast, away from the above mentioned infrared source.

\subsubsection{HH~1196 flow}

\begin{figure*}
\centering
\begin{tabular}{@{}ccc@{}}
\includegraphics[width=110pt]{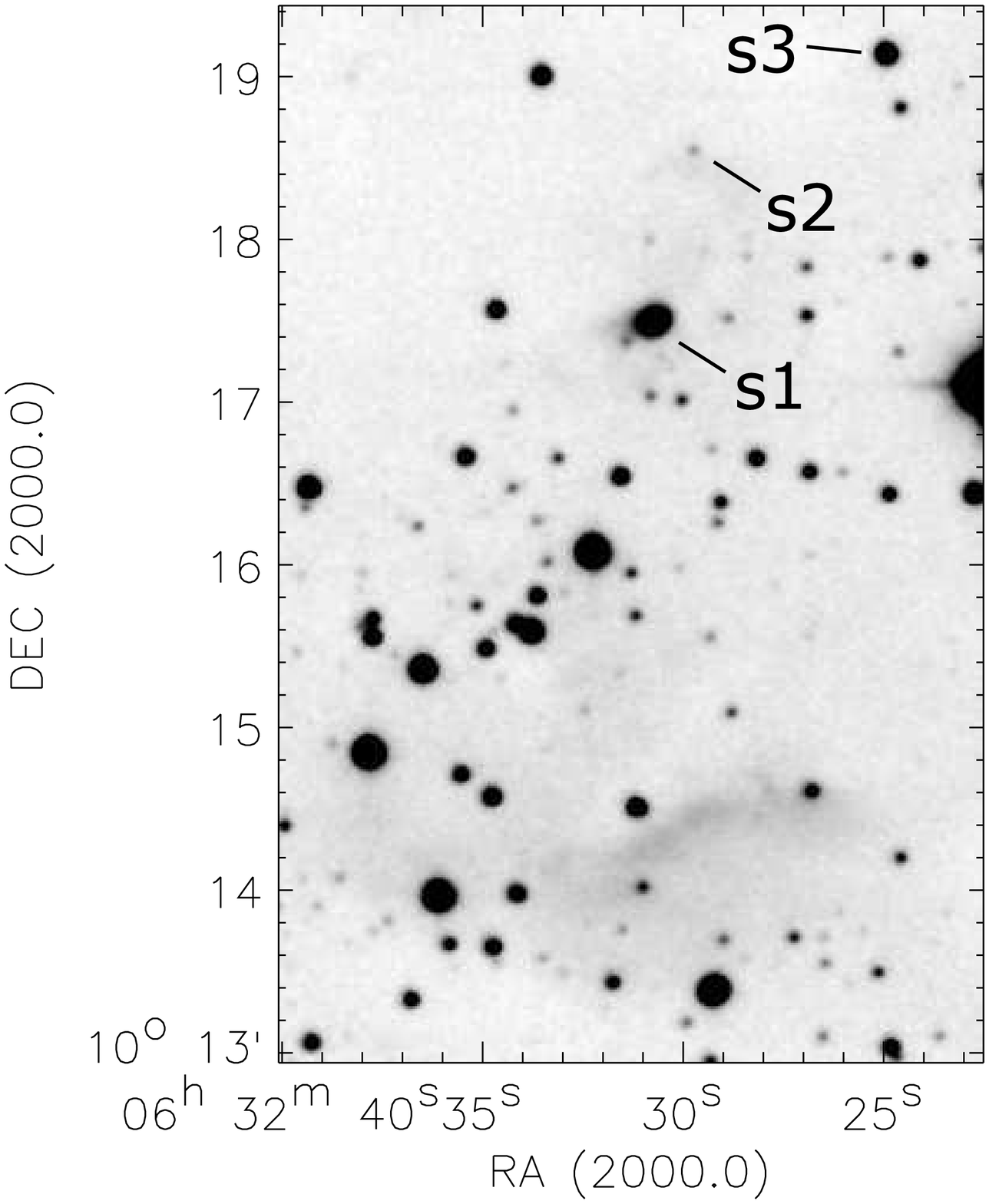}&
\includegraphics[width=110pt]{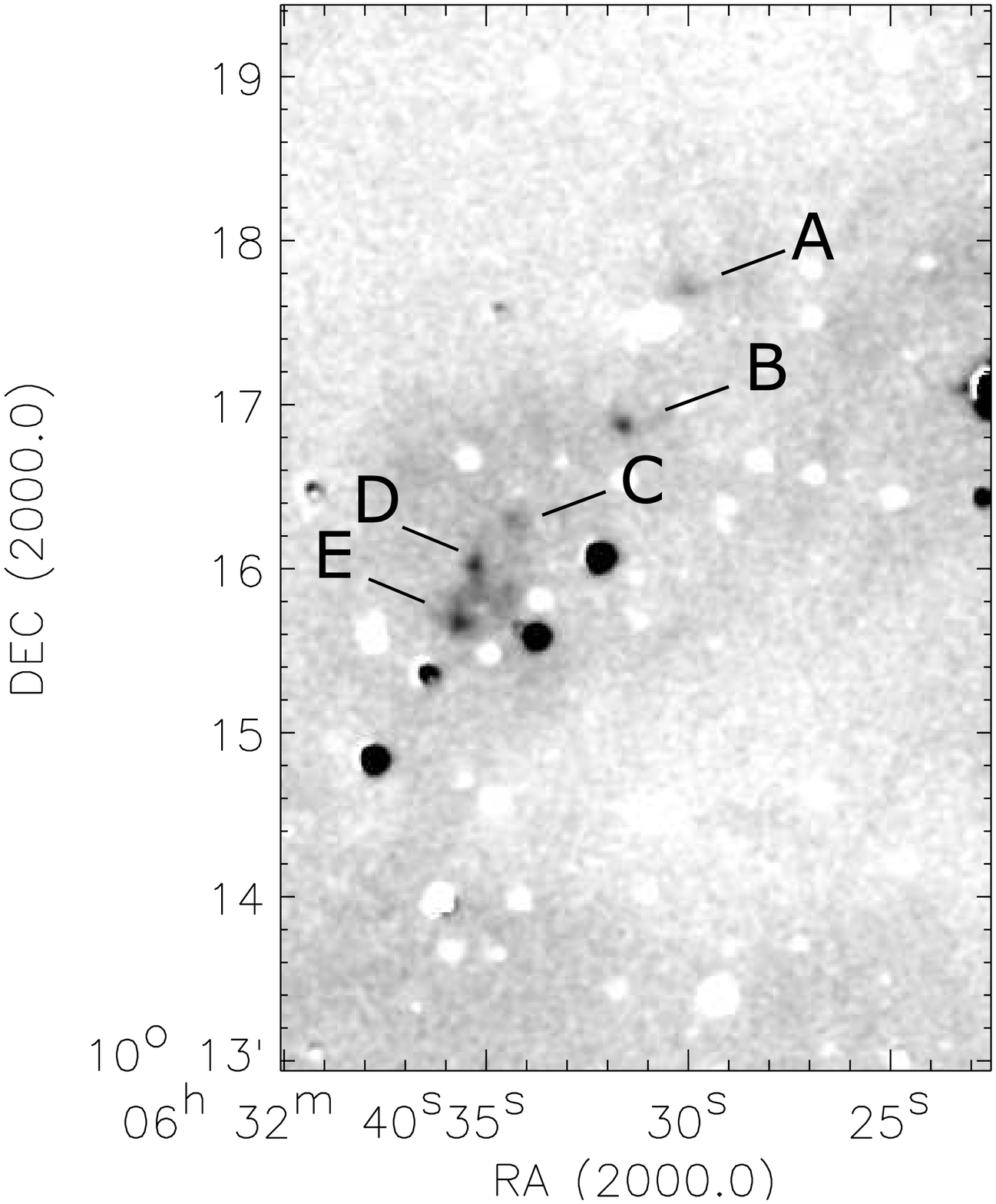}&
\includegraphics[width=108pt]{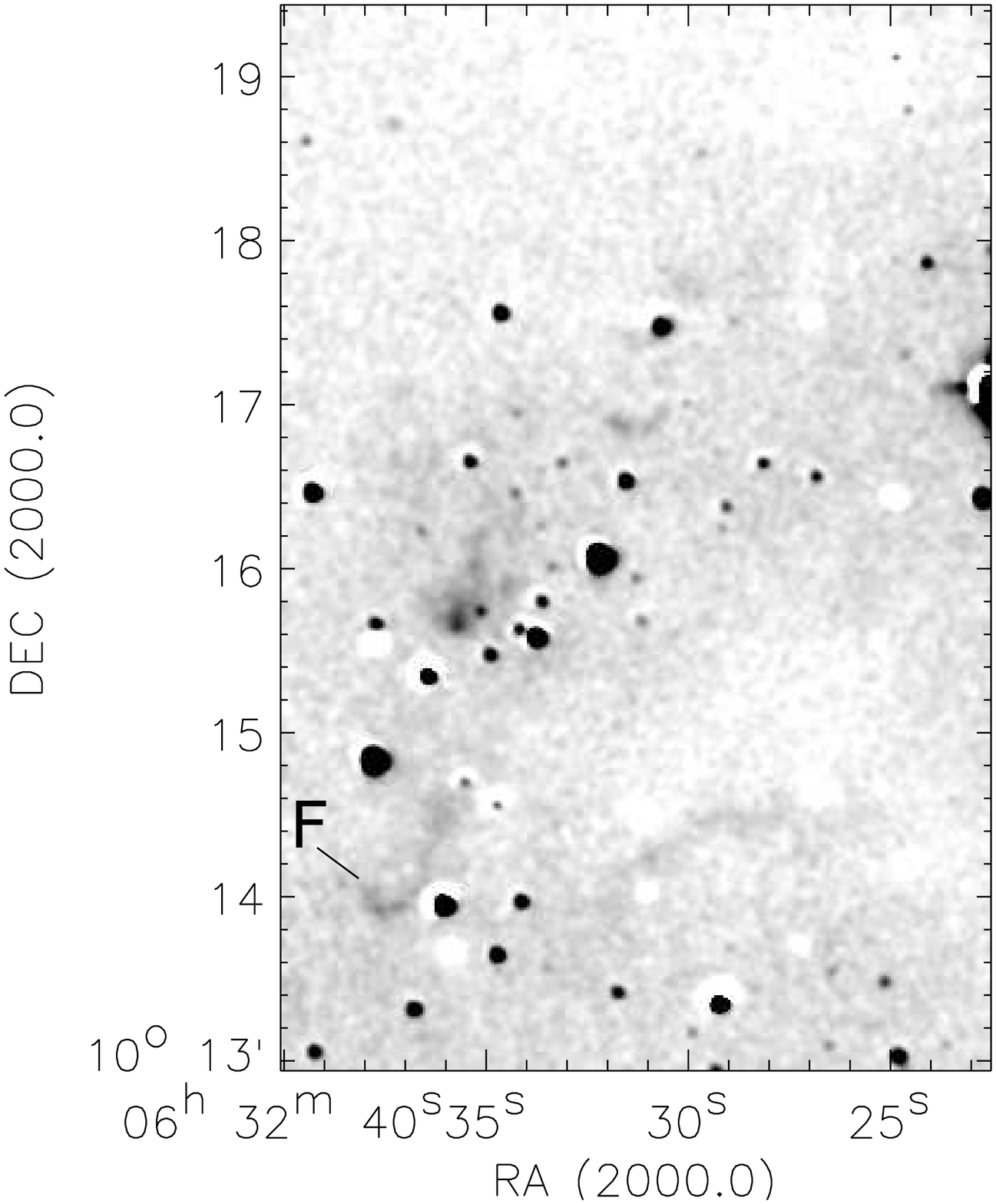}
\end{tabular}
\caption{Images of HH 1196 outflow system: continuum image (left panel),   continuum subtracted H$\alpha$ (central panel) and continuum subtracted [S \textsc{ii}] (right panel). By s1, s2 and s3 the 2MASS J06323159+1017352, IRAS~06297+1021 (E) and IRAS~06297+1021 (W) objects are marked respectively on the left panel.}
\label{HH1196}
\end{figure*}

This chain of HH objects contains at least six distinct knots with diffuse emission between them; its total length is about 4\arcmin\ (Fig.\,\ref{HH1196}). All knots are aligned along an axis with position angle of 155$\degr$. Knots A, B, E and F have bow-shape morphology with apexes pointing in S-SE direction. This fact confirms the impression of a single collimated outflow from a source, located in the northern direction. 

As can be seen from Fig.\,\ref{HH1196}, the relative intensities of the individual knots in H$\alpha$ and [S~\textsc{ii}] vary. Knots A, B, C and D are relatively brighter in H$\alpha,$ while knot E has the ratio of H$\alpha$ and [S~\textsc{ii}] near 1, and knot F is seen mainly in 
[S~\textsc{ii}].

Along the line connecting the HH~1196 knots, several objects, any of which  can be suspected as a possible source of this outflow, can be found. We discuss them in detail in next sections.

\begin{table*}
 \centering
 \begin{minipage}{190mm}
  \caption{The coordinates of nebulous stars in the observed field.}
  \begin{tabular}{@{}llllp{130pt}}
  \hline
   Name     &  RA(2000)    & Decl.(2000) & Other names & Notes  \\
 \hline
 IRAS 06277+1016  & 06 30 28.6 & $+$10 14 25.5 & 2MASS J06302857+1014236  &Far and mid IR source located in the apex of anon. cometary nebula \\
 HD 258686  &06 30 47.1   &   $+$10 03 46.4 & VdB 76 & Star with nearby reflection nebula \\
 2MASS J06305034+1037218 &   06 30 50.3 & $+$10 37 22.9 & Steine GN J0630.8+1037 & Star with tiny reflection nebula \\ 
 VY~Mon/G2 & 06 31 06.9 & $+$10 26 05.0 & HBC 522 & Small compact nebula, with no star inside\\
2MASS J06311641+1022326 & 06 31 16.4 & $+$10 22 32.6 & Ber 92 & Star with nearby reflection nebula \\
 LkH$\alpha$ 342   &   06 31 30.1 & $+$10 32 33.5 & HBC 204 & Emission-line star enveloped by a small reflection nebula \\
 HD 258973 &06 31 43.3 &  $+$10 20 20.9  &  VdB 79 & Star with nearby reflection nebula \\ 
 2MASS J06315782+1027360 & 06 31 57.8 & $+$10 27 35.6 &   RNO~72, HHL 43a & Brightest of three stars located in the apex of fan-like nebula RNO~72  \\
Gaia DR2 3327889958301335424 & 06 31 57.8 & $+$10 27 38.9 & RNO 72 & The star inside RNO~72 (``star 2'' in Fig.\ref{RNO72})
\\
2MASS J06315810+1027408 & 06 31 58.1 & $+$10 27 41.1 & RNO 72 & The star inside RNO~72 (``star 3'' in Fig.\ref{RNO72})
\\
 2MASS J06323082+1018396 & 06 32 30.8 & $+$10 18 39.6 & IRAS~06297+1021 (E) &Faint star, can be associated with very faint reflection nebula \\
 2MASS J06323159+1017352 & 06 32 31.6 & $+$10 17 35.2 & Petr 7  & Star in the head  of cometary nebula \\
  LkH$\alpha$ 216  &06 32 52.4  & $+$10 18 43.1  &  V490 Mon   & Variable star, surrounded by tiny reflection nebula  \\

\hline

\end{tabular}
\end{minipage}
\end{table*}

\subsection{Stars associated with reflection nebulae}

Being typical R-association, the investigated area contains significant amount of small reflection nebulae,  connected with one or more stars. Several of these nebulae besides  of being catalogued, never were studied in detail. Others are newly found objects, even not catalogued before. 

Besides, during searches of the possible exciting stars of HH flows we noted several candidates represented by interesting IR sources. We list the coordinates and short descriptions of all nebulous objects and possible outflow  sources in our field in Table 2 (excluding three brightest nebulae, identified in Fig.\ref{fig1}). Below they are discussed in more detail. 

\subsubsection{IRAS 06277+1016}

As was stated in Sec.3.1.1, HH~1203 flow is connected with a small cometary nebula. Near the apex of this reflection nebula a conspicuous stellar-like mid-IR source allWISE J063028.65+101425.4 can be found. It is located 5\arcsec\ to the east from the optical nebula and is well visible also in Spitzer GLIMPSE360 and Herschel PACS images.  We identify it with IRAS~06277+1016 source, though the later is displaced by more than 1\arcmin , because there are no other noticeable IR source in vicinity. Apparently, just this IR object represents the source of the collimated HH~1203 flow. Its position is shown by cross in Fig.\ref{HH1203Neb}. It is interesting that in 2MASS  images this source (2MASS~J06302857+1014236)  is faint, non-stellar and elongated
toward the optical reflection nebula. Such shift of the source position from the optically visible nebula is common  for the heavy 
embedded outflow sources, e.g. HH~83 or FS~Tau~B \citep{EM,MR}. 

\subsubsection{Steine J0630.8+1037}

This tiny reflection nebula (Fig.\ref{Steine}) is located in a small dark cloud in 10\arcmin\ to the N-NW from IC~446. It was listed among the results of a large search of new galactic open clusters \citep{kronberger}, as Steine J0630.8+1037. The nearby (in 30\arcsec\ distance) source IRAS 06281+1039 can be related to this object. The nebula is connected with a red star (2MASS J06305034+1037218), which quite recently faded by one magnitude, according to data from Gaia transient survey (Gaia19drn). Without any doubts this object represents an active YSO; however, search in its vicinity did not reveal any HH objects.

\subsubsection{IC~446/VY~Mon\ field}

This field actually includes two reflection nebulae (see Fig.\ref{VYMon}). IC~446 itself consists of a blue nebula, bright in the optical range and illuminated by B2.5V TYC~737-255-1 (or IC~446 No. 1) star, and a rather bright and very red nebula, well seen even in FIR range. Its excitation source, as was shown by \citet{casey}, is VY~Mon. Here we want to draw attention to a small, but compact and bright nebula,  for the first time described by \citet{maffei}, who also  mentioned its probable variability. This nebula is designated as VY~Mon/G2 in a fundamental survey of \citet{ck}. It is very close to VY~Mon; however, according to the same survey, its spectral type is A0, while VY~Mon itself is classified as O9. Nevertheless, more recent data give for the spectral type of VY~Mon estimates between B8 and A5. Thus, it seems that this nebula indeed can be a dense knot of dust, illuminated by VY~Mon. This object as HBC~522 was also included in the catalog of PMS stars \citep{HBC}. In any case, both our observation and 2MASS images as well do not show any stellar source inside VY~Mon/G2.    

\subsubsection{RNO~72}

This nebula was included in the early lists of red nebulous objects in dark clouds \citep{RNO,HHL43}. No further studies were performed. However, besides of our finding of a nearby HH~1197 knot,  one should mention an  interesting structure of this object as a whole.

\begin{figure*}
\centering
\begin{tabular}{@{}ccc@{}}
\includegraphics[width=130pt]{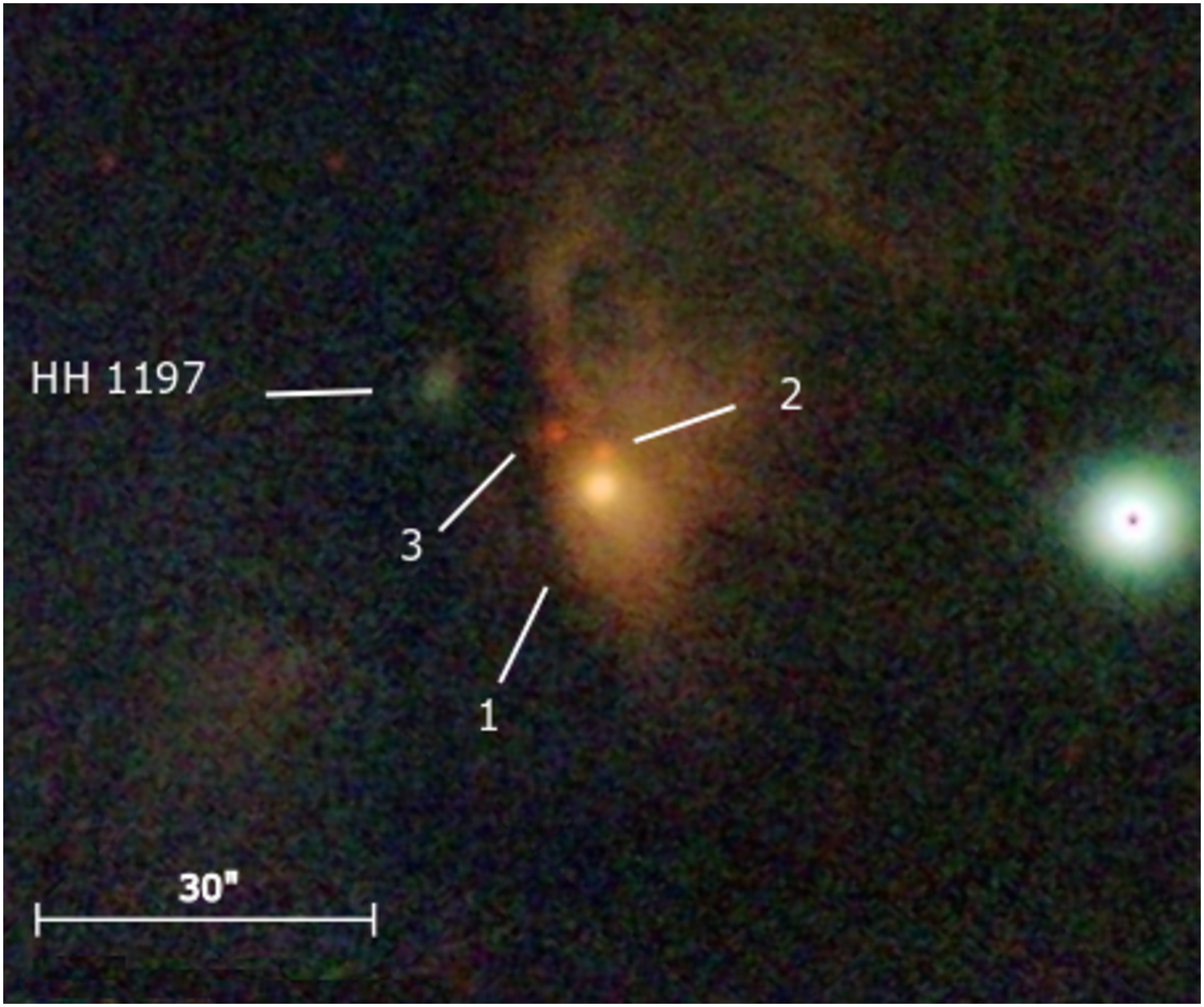}&
\includegraphics[width=138pt]{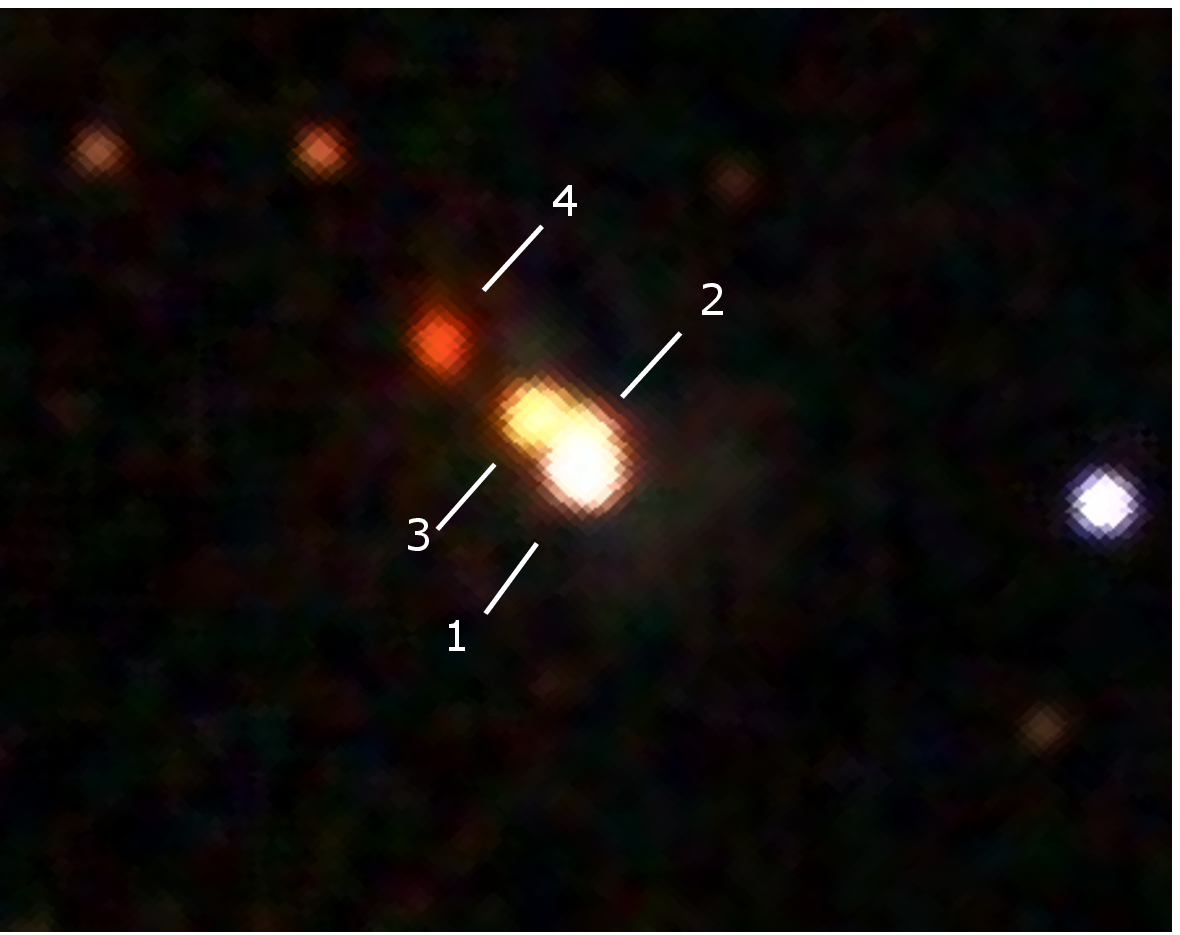}&
\includegraphics[width=142pt]{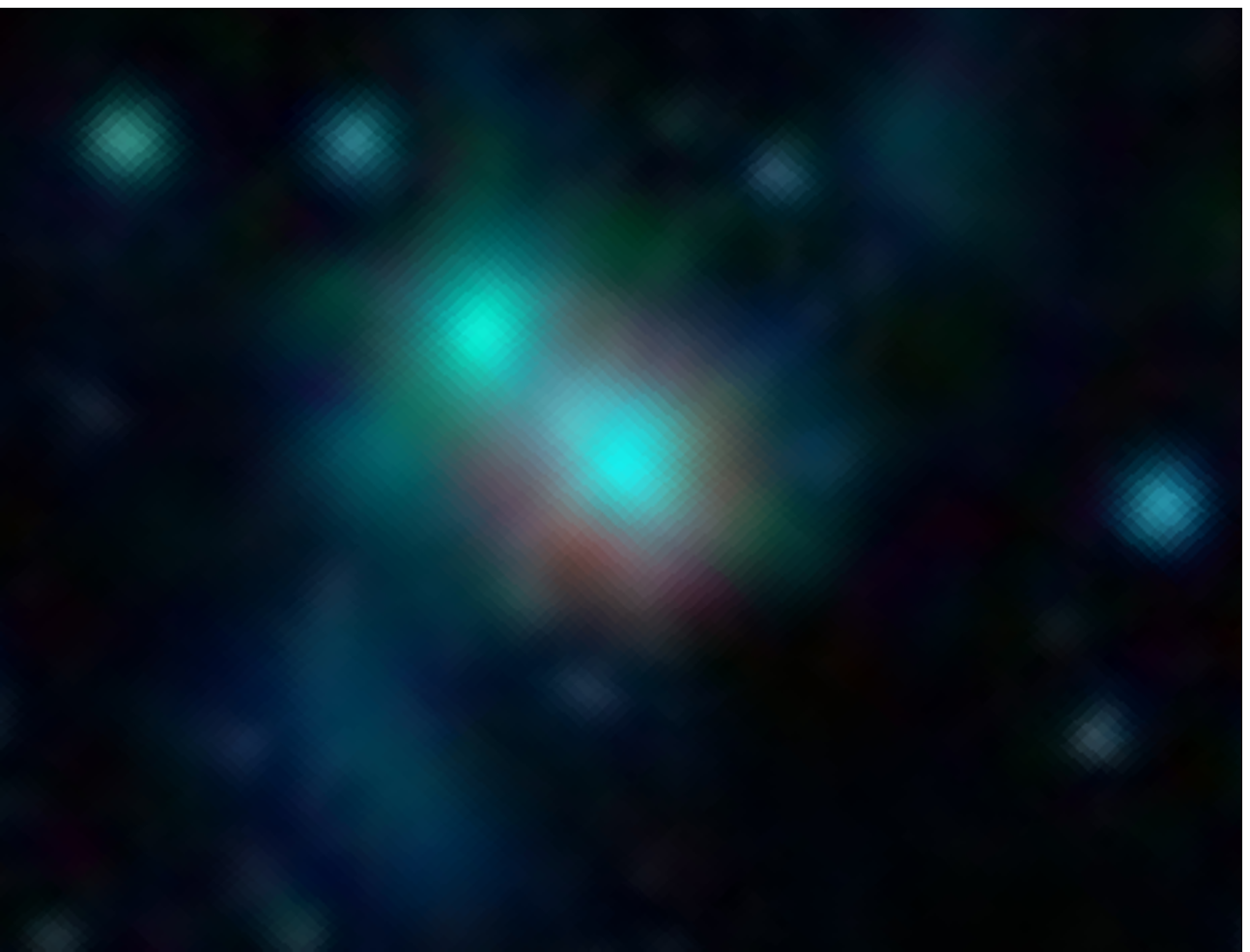}
\end{tabular}
  \caption{Objects in the field of RNO 72 nebula: color images from  PanSTARRS survey (left panel), 2MASS survey (central panel) and Spitzer GLIMPSE360 survey (right panel). Numbers denote the following stellar sources: 1 - 2MASS J06315782+1027360, 3 - 2MASS J06315810+1027408, 4 - 2MASS J06315870+1027474 (IRAS 06292+1029). Star 2 was not resolved in 2MASS, but has a Gaia DR2 number (see Table 2 and Table 4). Color coding for PanSTARRS: blue (g-band), green (r-band), red (i-band). Color coding for GLIMPSE360: blue (3.6 $\mu$m), green (4.5 $\mu$m), red (12 $\mu$m).  }
        \label{RNO72}
\end{figure*}  

\begin{figure}
        \includegraphics[width=200pt]{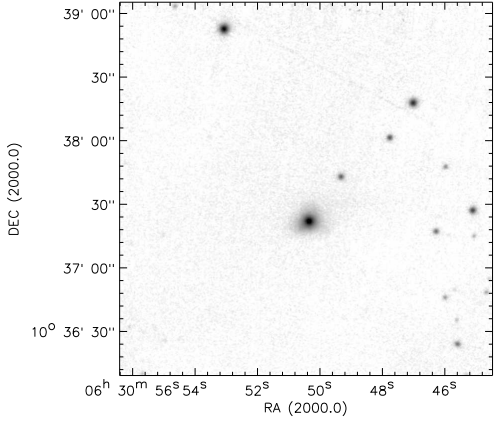}
        \caption{Reflection nebula Steine J0630.8+1037 in PanSTARRS r image}
        \label{Steine}
\end{figure}

\begin{figure}
        \includegraphics[width=230pt]{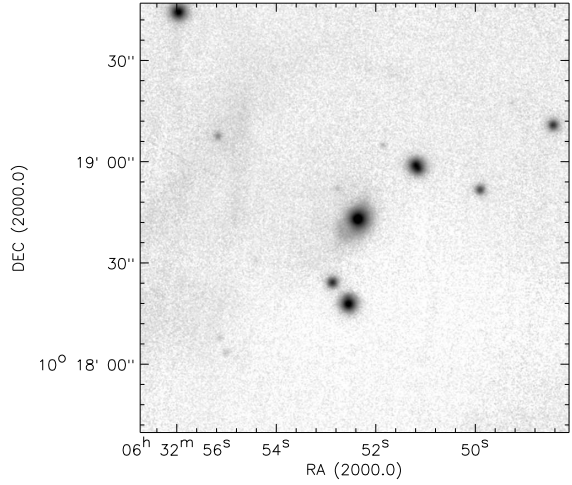}
        \caption{LkH$\alpha$ 216 and its surrounding nebula in PanSTARRS r image.}
        \label{Lk216}
\end{figure}       

\begin{figure}
        \includegraphics[width=230pt]{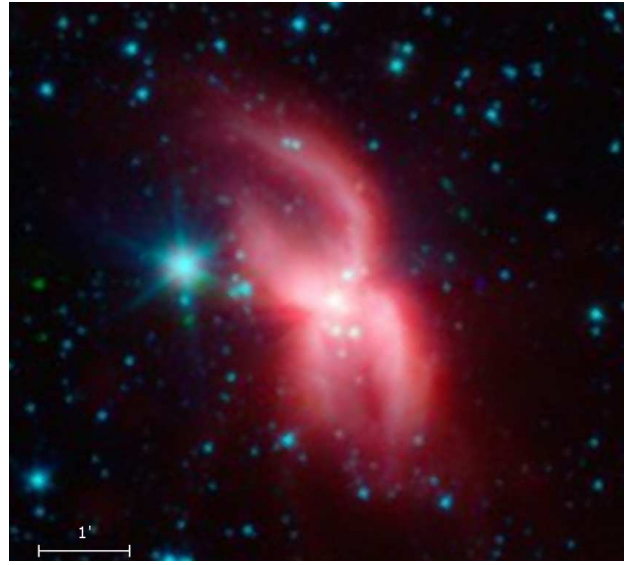}
        \caption{NGC 2245 nebula in Spitzer GLIMPSE360 survey. The bright star in the center is  LkH$\alpha$ 215.}
        \label{NGC2245}
\end{figure}       

\begin{figure}
        \includegraphics[width=230pt]{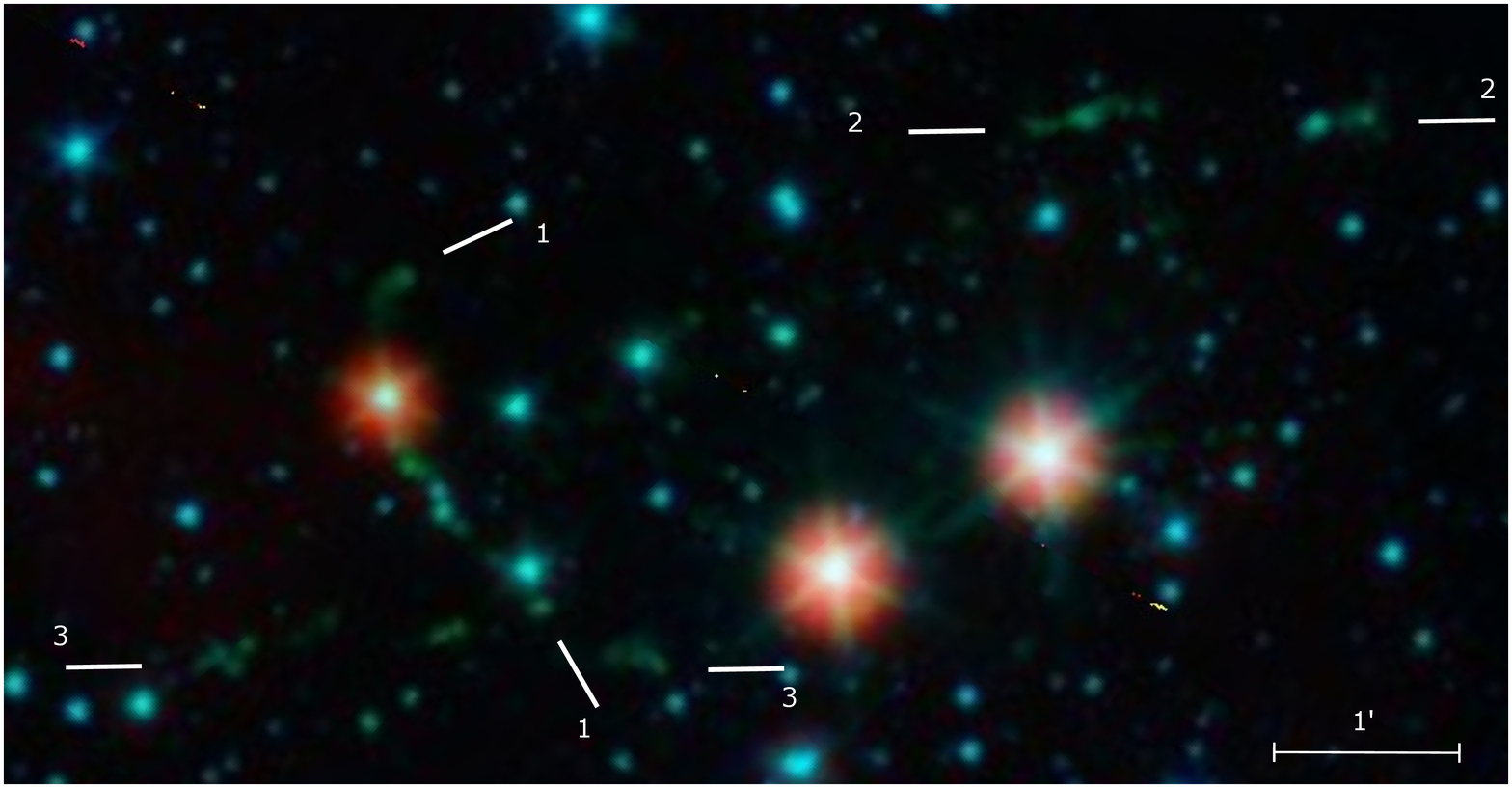}
        \caption{Three probable  H$_{2}$ flows, seen in Spitzer GLIMPSE360 survey. Two bright red objects near the center are  IRAS~06297+1021 (E) and IRAS~06297+1021 (W) sources.}
        \label{h2flows}
\end{figure}     
The brightest part of this nebula has roughly fan-like shape, while filaments and other details can be traced up to 45-50\arcsec\ from the center (Fig.\ref{RNO72}, left). They can be seen also in Spitzer GLIMPSE360 color image (Fig.\ref{RNO72}, right). Though HH~1197 knot is clearly seen in PanSTARRS images (perhaps due to the continuum component in its spectrum), there are virtually no traces of the above mentioned 2MASS source J06315870+1027474 in the optical range. Besides, there is a close group (\textless 6.5\arcsec \ in length) of three stars, embedded in RNO 72 itself; the brighter one (2MASS J06315782+1027360) is the probable illuminator of the optical nebula (Fig.\ref{RNO72}, left and center).     All three stars are prominent sources in near IR and remain noticeably bright in longer wavelengths. Thus, one can conclude that in the center of RNO 72 exists a whole small cluster of PMS stars.      

\subsubsection{IR objects near HH~1196 flow}

 As we mentioned above, there are three possible sources of HH~1196 outflow. The star associated with a small comet-like reflection nebula (No.7 in the list of \citealt{P85}, marked as s1 on Fig.\,\ref{HH1196}) is located just between knots A and B; however, the morphology of bow-shaped knot A suggests the more distant north-western source location. Thus, as other and more probable sources of this outflow  the YSOs IRAS~06297+1021 (E) (2MASS J06323082+1018396)   and IRAS~06297+1021 (W) (2MASS J06322611+1019184) (s2  and s3 on Fig.\,\ref{HH1196}), can be considered. Especially interesting is the IRAS~06297+1021 (W)  source, which is rather bright in mid-IR range and has strong  emission lines in IR spectrum; some features in its spectrum  also are in common with FU Ori-like objects \citep{Connelley2010}. In any case,  these features indicate the existence of high-rate outflow activity in this YSO. Besides,  IRAS~06297+1021 (W) is located close to the direct line, connecting all knots of HH 1196 outflow. 
If this object indeed is its source, the total length of HH~1196 outflow system is about 6\arcmin, which on the distance of 800 pc will be equivalent to about 1.4 pc, making it a parsec-sized outflow.

It is worth to mention that IRAS~06297+1021 (W), as well as IRAS~06297+1021 (E)
are well visible in the optical range, but there are no data about their optical spectra. HH 1196 flow definitely deserves further study. 

\subsubsection{LkH$\alpha$ 216, NGC 2247 and NGC 2245}

The emission-line star LkH$\alpha$ 216 (V490 Mon) was found by \citet{Herbig1960}  in the vicinities of NGC 2247 nebula in his classic paper about HAeBe stars, and its variability was noted by \citet{maffei}. Strangely, it remains very  little studied up to now, and even its spectral type never have been described; it is not included in HBC catalog.
Our data as well as DSS-2 and PanSTARRS
images show that LkH$\alpha$ 216 is surrounded by a small elliptical nebula (Fig.\ref{Lk216}).

The attention to the bright and large reflection nebulae NGC 2245 and NGC 2247 was drawn, when \citet{Herbig1960} pointed that both nebulae are  illuminated by young HAeBe stars LkH$\alpha$ 215 and HD 259431 respectively. In the same work it was also suggested that NGC 2245 can have biconical morphology, though it is not very well seen in the optical images. However, in Spitzer GLIMPSE360 survey   the nearly perfect X-shaped nebular structure is obvious (Fig.\ref{NGC2245}).        

\subsection{Probable molecular hydrogen flows}

Analysing the mid-IR images of our field from Spitzer GLIMPSE360 and WISE surveys, we noted several knots arranged in chains and visible mainly in 4.5 $\mu$m bands both of  Spitzer and WISE, but not detected in 3.6 $\mu$m. All of them are located in a very opaque area to the west from NGC 2247 nebula, around  IRAS~06297+1021 (W) and (E) sources.
Very probably they represent H$_2$\ flows from deeply embedded PMS objects. They are shown in Fig.\ref{h2flows} and the coordinates of their approximate centers  are listed in Table 3. For the  most conspicuous of them, probably connected with WISE J063240.86+101937.4, we give coordinates of this IR-source.  Of course, the full length and structure of these flows can be understood only after further observations in 2.12 $\mu$m  line, which are planned\footnote{Our assumption that these chain-like structures can be molecular hydrogen flows, was fully confirmed by new observations. They will be described in the forthcoming paper.}.

\begin{table}
 \centering
 \begin{minipage}{110mm}
  \caption{The coordinates of probable H$_{2}$ flows in the Mon~R1 field.}
  \begin{tabular}{llll@{}}
  \hline
   No.     &  RA(2000)    & Decl.(2000) & Notes  \\
 \hline
Flow 1 & 06 32 40.9 & +10 19 37.3 &   IR-source with narrow flow\\
Flow 2 & 06 32 23.9 & +10 21 11 & At least two elongated knots\\
Flow 3 & 06 32 39.8 & +10 18 19 & At leat four knots \\
 \hline

\end{tabular}
\end{minipage}
\end{table}

\section{Discussion}

 The presented above results confirm the active star formation in Mon R1. Recent surveys and catalogs help to obtain more precise characteristics for this complex. We tried to obtain the new estimate of Mon R1 distance and to enlarge and clarify the list of its members with the aid of the recent catalog of \citet{BJ}, based on Gaia DR2 results.
We assumed the data for the bright and definitely connected with the dark cloud LkH$\alpha$ 215 and HD 259431
stars (703 and 711 pc respectively) as a reference distance. Then we compared it with the parallaxes of other bright (V $<$ 11) stars of B and A types from the updated list of \citet{Herbst} and got 716 $\pm$ 47 pc for mean distance of Mon~R1. It is worth to mention that the supposed members VdB~71 and VdB~81 \citep[see][]{Racine}  were rightly excluded from the updated list, because Gaia parallaxes indeed confirm that they are background  and    
foreground objects. Also not quite clear case is HD~258686 (or VdB~76), where component A (well-known magnetic CP star) of this close (1.7\arcsec) binary has, according to Gaia DR2, distance 591 pc, while component B -- 675 pc. Thus, its duplicity could be an optical one.
        
  Comparing the other objects from the list of Mon~R1 members with our Table 2, we see
that they match almost completely, including RNO~72 and Petr 7 nebulae, which were independently found by  \citet{Herbst}.
In Table 4 we collected
all stars from both lists, matched with our field,   and their Gaia distances (if such exist).

A special case, which was not included in Table 4, is a small cluster of late-type emission-line stars around VY~Mon. First this grouping was described and studied by \citet{ck}, and then by \citet{Herbst}  and \citet{Habart}. VY Mon is considered by \citet{Herbst} as the most luminous object in Mon R1. Its  parallax was not correctly measured by Gaia, but the distance of the nearby TYC 737-255-1, which illuminates  another part of the same nebula, is 641 pc.  Though the analysis and discussion of VY Mon cluster is beyond the scope of this paper, we found that the good quality Gaia parallaxes were obtained for all its stars, excluding V540 Mon (or CoKu VY Mon G4, which turned out to be a close double). Average distance of six stars of this group equaled 652 $\pm$ 27 pc. The perceptible difference of distances lets to suspect that we see here several overlapped clouds, close to each other.

Another case for the separate study present several  emission-line stars, mainly near LkH$\alpha$~215 and HD~259431, discovered by \citet{Herbig1960}. Some of them probably should belong to Mon~R1, but the data are scarce. For example, LkH$\alpha$~216 with 563 pc distance  is probably  a foreground object.

  We assume 715 $\pm$ 50 pc as a good estimate for the distance of Mon~R1 association. This, in turn, puts it on the same distance as Mon~OB1 \citep{Zucker}.

\begin{figure*}
        \includegraphics[width=400pt]{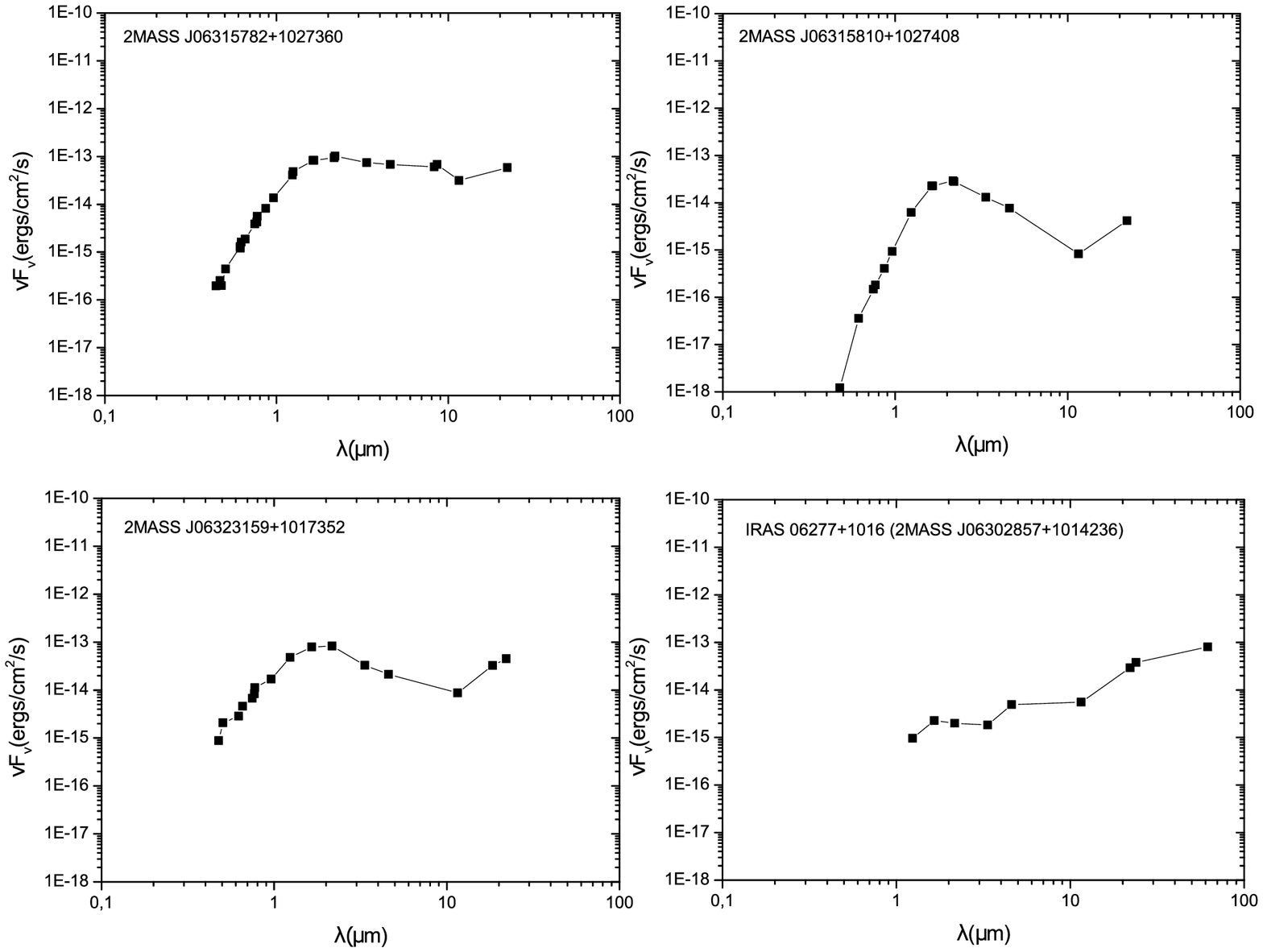}
        \caption{SEDs for the several stars from Table 4.}
        \label{SEDs_a}
\end{figure*}

\begin{figure*}
        \includegraphics[width=400pt]{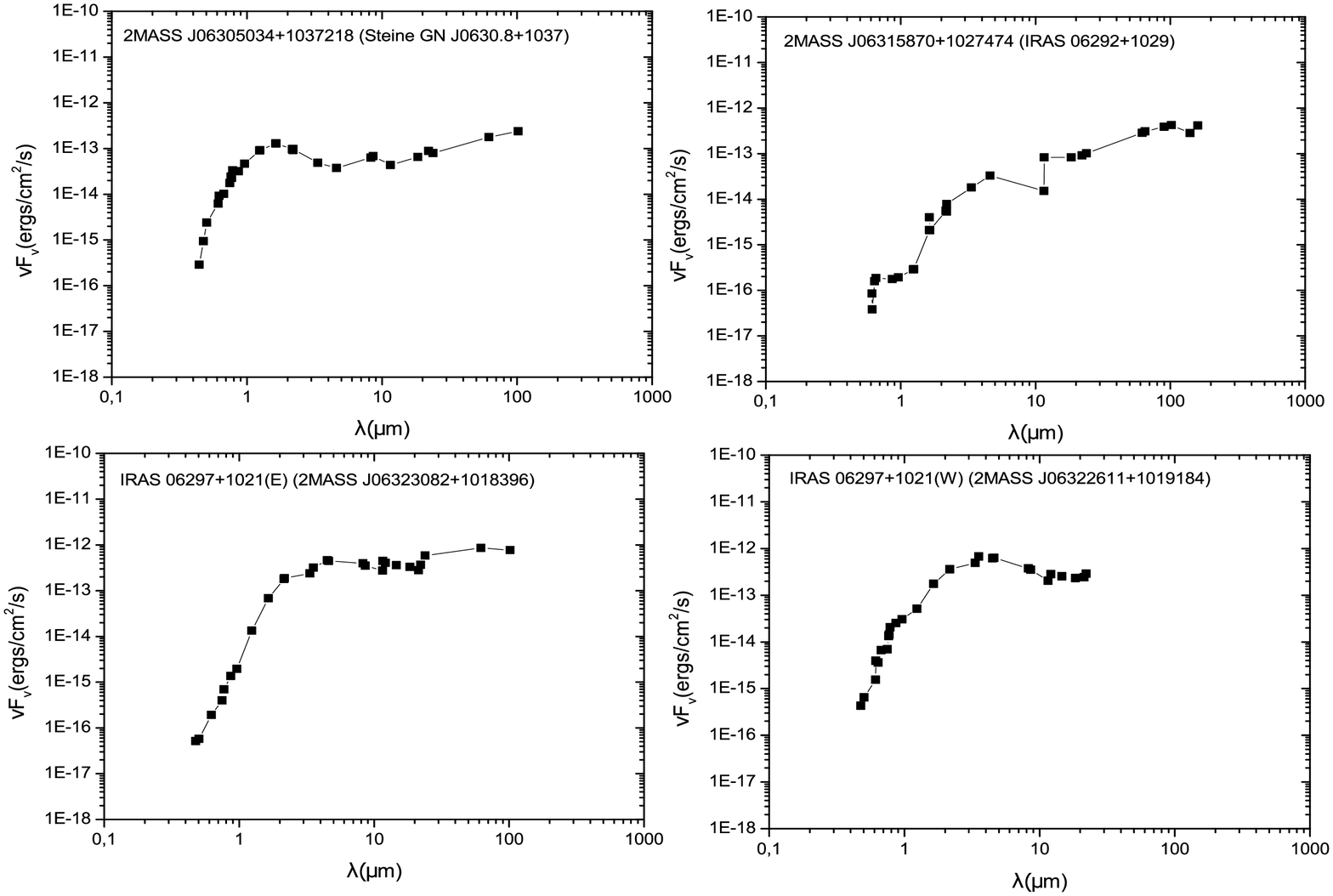}
        \caption{SEDs for the several stars from Table 4.}
        \label{SEDs_b}
\end{figure*}

With the aid of \textit{Vizier photometry viewer} we built spectral energy distributions (SED)
for the main part of all stellar sources listed in Table 4, with exception of the brightest, well-studied stars and the described above groupings.
2MASS J06311641+1022326, HD~258973,  LkH$\alpha$~342 have no perceptible IR excess.
All others are
prominent IR emitters. Their SEDs are shown in Fig.\ref{SEDs_a} and Fig.\ref{SEDs_b}.

It is obvious that all these stars are surrounded by dusty envelopes and/or disks. On the other hand, it is remarkable that nearly all of them are visible in the optical range. Thus, foreground extinction cannot be too high.

Several of these SEDs should be considered with caution.
It was not possible to obtain the separate photometry for RNO 72 stars 1
and 2; therefore the shown SED relates to both stars together.
In the case of IRAS~06297+1021 (W) and IRAS~06297+1021 (E) it was not obvious how to consider the data of IRAS photometry, because the  IRAS source coordinates are exactly between the both stars.

At least three stars from Table 4 are considered as HAeBe stars, i.e. intermediate mass objects. We checked their luminosities in the new catalogue, based on the Gaia DR2 data \citep{Vioque}. For LkH$\alpha$ 215  and HD 259431 we inserted these estimates
in the Table 4. However, for VY Mon their estimate is quite improbable due to wrong (negative) Gaia parallax. Thus, for this star we tried several other approaches. The direct integration of VY Mon SED gives for its luminosity 640 L$_{\sun}$ value, while the estimate from the relation, based only on IRAS fluxes \citep{CRT}, is merely 325  L$_{\sun}$ (but it does not take into account that the significant part of VY Mon flux is emitted in the $< 10 \mu$m range). Also it is possible to compute extinction-free M$_V$ from the photometric data of \citet{Herbst}, which leads to 524  L$_{\sun}$.
Since sub-mm radiation of VY Mon should be significant \citep[see][]{casey}, 800-900  L$_{\sun}$ can be a good estimate for its total bolometric luminosity. In any case VY Mon is one of the most massive and luminous members of Mon R1.
The analysis of nearby TYC 737-255-1 is beyond the scope of this paper, but it is obvious that the luminosity of this star
was greatly overestimated before, since its distance is two times lower than previously assumed.

For the stars with SEDs, shown in Fig.\ref{SEDs_a} and Fig.\ref{SEDs_b}, bolometric luminosities were estimated      by SED integration.
As can be seen, all objects, measured up to 100 $\mu$m or more, have flat or even rising spectra. However, application of the abovementioned relation from \citet{CRT} to these stars gives almost exactly matching luminosity values (note that the luminosity of IRAS~06297+1021 (E), given by \citet{Connelley2010}, is computed for exaggerated 900 pc distance).  SEDs of other stars without far-infrared data have similar shapes. Of course, their luminosities are underestimated. Since no IRAS data exist, we tried to compute the bolometric corrections using the approach suggested by \citet{Cohen73}.
This correction, if applied, increases our estimates only to 1-3  L$_{\sun}$, not changing them drastically.   We assume that even if our estimates constitute only lower limits for stellar luminosities, they are not too understated.

\begin{figure}
        \includegraphics[width=230pt]{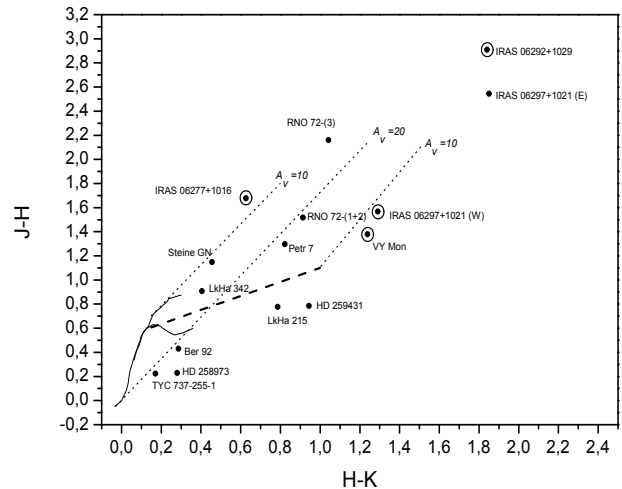}
        \caption{J-H/H-K diagram for all stars from Table 4. Probable sources of HH flows are marked by circles. Main sequence and giant branch are shown by solid lines. Dashed line shows T Tauri type stars locus.}
        \label{JHK}
\end{figure}

For the better understanding of the stellar content of Mon~R1 we drew a $H-K/J-H$ diagram \citep{BB} for the all stars from Table 4. It is shown in Fig.~\ref{JHK}. Its analysis allows to come to the following conclusions. All stars without IR excess have rather low extinction and fall to the beginning of main sequence. HAeBe stars
can be found near the so-called T Tau stars locus \citep{MCH}. This can be interpreted as an evidence of the  low foreground extinction. Other sources are much more reddened (A$_V$ from 5 to 20). After taking it into account they also could be brought into vicinities of T Tau stars locus. As can be seen from Fig.~\ref{JHK}, the most extincted objects are IRAS~06297+1021 (E) and IRAS~06292+1029, source of HH~1197. Such large values  of extinction, however, contradict the fact that nearly all these objects are detected in the optical range. One can explain it by existence of circumstellar dust disks, which radiate in IR range, while certain amount of stellar light escapes through the central parts of discs with low optical density and reaches us in the visible range.

We come to conclusion that our observations revealed at least three stars with L $\geqslant$\ 10 L$_{\sun}$ in the studied part of Mon R1, besides of three more
luminous HAeBe stars. At least three of these six are probable sources
of HH outflows, and IRAS~06297+1021 (W) is even a suspected FU Ori type object. The newly found sources definitely deserve detailed studies. The number of such objects in Mon R1 eventually
can be further increased when it will become possible to identify sources of molecular hydrogen outflows (see Sec. 3.3).

\section{Conclusion}
 
The discovery of new HH outflows in Mon R1 association shows that the star formation in this area is  more active than appeared before and continues in the present time. In comparison with the nearby
Mon OB1 association we see that even if Mon R1 does not produce high-mass stars, it contains a significant amount of active stars of low and intermediate mass, among which can exist even such rare objects as FUor and EXor like stars. In fact, Mon OB1 and Mon R1  probably are the parts of one giant molecular cloud, and the differences in the rate of star formation and in the luminosity function of YSOs can be important for the better understanding of the physical processes that govern the
clouds fragmentation and the formation of massive
stars and clusters of YSOs (see also \citet{Bhadari}, \citet{Montillaud} and references therein).     

Taking into account that nearly all sources, found by us, are visible in the optical range, they can  become a good follow-up targets for the spectral studies with large telescopes. Besides, as was stated above,  HH outflows (especially the giant ones) reflect the history of outburst activity; thus, their long-slit spectroscopy will allow to study the ejection kinematics. Such observations are already planned by our group.

  \begin{table*}
 \centering
% \begin{minipage}{190mm}
  \caption{Stellar members of Mon~R1 association in the studied field}
  \begin{tabular}{@{}llllp{92pt}l}
  \hline
   Name     &  D (pc)$^a$  & IR-excess & HH objects & Nebula & L$_{bol}$ (L$_{\sun}$) \\
 \hline
 IRAS 06277+1016 (2MASS J06302857+1014236)  &  & yes & HH 1203  & Anon &1.6 \\

 2MASS J06305034+1037218 (IRAS 06281+1039?) &   710 & yes &  & Steine GN J0630.8+1037 & 6--8$^b$ \\ 
 TYC 737-255-1 & 641 & no & & IC 446 & $<$900?\\
 VY Mon & & yes & HH 1202 & IC 446 & 800--900$^b$\\
 2MASS J06311641+1022326 & 664 & no & & Ber 92 & 3 \\
 
 LkH$\alpha$ 342 (HBC 204)   &  & no & HH 1201 ? & & 2.8\\
 HD 258973 &750 &  no  & & VdB 79 & 75$^c$\\ 
 2MASS J06315782+1027360 &  & yes &  &  RNO 72 star 1 &\rdelim\}{2}{1cm}[3.3]   \\
 Gaia DR2 3327889958301335424 & & yes & & RNO 72 star 2 & \\
 2MASS J06315810+1027408 & & yes & & RNO 72 star 3 & 0.5\\
 2MASS J06315870+1027474 (IRAS 06292+1029) & & yes & HH 1197 & & 10--11$^b$ \\
 IRAS~06297+1021 (W) (2MASS J06322611+1019184) & 759 & yes & HH 1196 & & 20  \\
 IRAS~06297+1021 (E) (2MASS J06323082+1018396) &  & yes & ?  & Anon & 30--35$^b$ \\
 2MASS J06323159+1017352 & & yes &   & Petr 7 & 2  \\
 LkH$\alpha$ 215 & 703 & yes & & NGC 2245 & 372$^d$ \\
 HD 259431 & 711 & yes & & NGC 2247  & 933$^d$  \\

\hline

\end{tabular}
\begin{flushleft}
\textit{Notes}: \\
$^{(a)}$  Determined from Gaia parallaxes \citep{BJ}. \\
$^{(b)}$ Estimates based on several approaches (see text). \\
$^{(c)}$ On the base of photometry from \citet{Herbst}. \\
$^{(d)}$ From the catalogue of \citet{Vioque}. \\
All other luminosities are determined solely by SED integration.
\end{flushleft}
%\end{minipage}

\end{table*}

 Besides of the all presented above results, this work demonstrates that with 1 m Schmidt telescope of Byurakan observatory, by means of which such well-known surveys of active galaxies, as First Byurakan Survey and Second Byurakan Survey, were conducted several decades ago,  still important achievements can be made.

\section*{Acknowledgments}

We thank Prof. Bo Reipurth for providing the numbers for new HH objects and  for very helpful suggestions. We thank referees, whose comments were of great help to improve the paper.  
This work was supported by the RA State Committee of Science, in the frames of the research project number 18T-1C-329.
This research has made extensive use of Aladin sky atlas, VizieR catalogue access tool and SIMBAD database,
which are developed and operated at CDS, Strasbourg Observatory, France.
The Spitzer GLIMPSE360 images are shown by courtesy NASA/JPL-Caltech. This publication makes use of data products from
the Two Micron All Sky Survey, which is a joint project of the University of Massachusetts and the
Infrared Processing and Analysis Center/California Institute of Technology, funded by the National
Aeronautics and Space Administration and the National Science Foundation. This publication makes use
of data products from the Wide-field Infrared Survey Explorer, which is a joint project of the
University of California, Los Angeles, and the Jet Propulsion Laboratory/California Institute of
Technology, funded by the National Aeronautics and Space Administration. This  work  is based  in  part  on  observations  made  with  the Spitzer SpaceTelescope, which is operated by the Jet Propulsion Laboratory,California Institute of Technology under a contract with NASA. The Pan-STARRS1 Surveys (PS1) and the PS1 public science archive have been made possible through contributions by the Institute for Astronomy, the University of Hawaii, the Pan-STARRS Project Office, the Max-Planck Society and its participating institutes, the Max Planck Institute for Astronomy, Heidelberg and the Max Planck Institute for Extraterrestrial Physics, Garching, The Johns Hopkins University, Durham University, the University of Edinburgh, the Queen's University Belfast, the Harvard-Smithsonian Center for Astrophysics, the Las Cumbres Observatory Global Telescope Network Incorporated, the National Central University of Taiwan, the Space Telescope Science Institute, the National Aeronautics and Space Administration under Grant No. NNX08AR22G issued through the Planetary Science Division of the NASA Science Mission Directorate, the National Science Foundation Grant No. AST-1238877, the University of Maryland, Eotvos Lorand University (ELTE), the Los Alamos National Laboratory, and the Gordon and Betty Moore Foundation. This publication has made
use of data from the European Space Agency (ESA) mission
{\it Gaia} (\url{https://www.cosmos.esa.int/gaia}), processed by the {\it Gaia}
Data Processing and Analysis Consortium (DPAC,
\url{https://www.cosmos.esa.int/web/gaia/dpac/consortium}). Funding for the DPAC
has been provided by national institutions, in particular the institutions
participating in the {\it Gaia} Multilateral Agreement.

\section*{Data availability}

The data underlying this article will be shared on reasonable request to the corresponding author.

\label{lastpage}

\end{document}